\documentclass[12pt]{article}

\usepackage{cite}
\usepackage{epsfig}
\usepackage{amsmath}
\usepackage{amssymb}
\usepackage{authblk}
\usepackage{url}
\usepackage[T1]{fontenc}
\usepackage[latin1]{inputenc}
\usepackage{textcomp}

\usepackage{array}
\usepackage{multirow}
\usepackage{slashed}
\usepackage{float}
\usepackage{xcolor}

\def\mathswitch#1{\relax\ifmmode#1\else$#1$\fi}
\def\mathswitchr#1{\relax\ifmmode{\mathrm{#1}}\else$\mathrm{#1}$\fi}
\newcommand{\PW}{\mathswitchr W}
\newcommand{\PZ}{\mathswitchr Z}
\newcommand{\PH}{\mathswitchr H}

\newcommand{\Pb}{\mathswitchr b}
\newcommand{\Pt}{\mathswitchr t}

\newcommand{\MW}{\mathswitch {M_\PW}}
\newcommand{\MZ}{\mathswitch {M_\PZ}}
\newcommand{\GZ}{\mathswitch {\Gamma_\PZ}}
\newcommand{\GW}{\mathswitch {\Gamma_\PW}}
\newcommand{\MH}{\mathswitch {M_\PH}}

\newcommand{\mb}{\mathswitch {m_\Pb}}
\newcommand{\mt}{\mathswitch {m_\Pt}}

\newcommand{\scrs}{\scriptscriptstyle}
\newcommand{\sw}{\mathswitch {s_{\scrs\PW}}}
\newcommand{\cw}{\mathswitch {c_{\scrs\PW}}}

\newcommand{\mz}{\mathswitch {M_\PZ^{\mathrm exp}}}

\newcommand{\gz}{\mathswitch {\Gamma_\PZ^\mathrm{exp}}}
\newcommand{\as}{\alpha_{\mathrm s}}
\newcommand{\at}{\alpha_\Pt}
\newcommand{\seff}[1]{\sin^2\theta_{\rm eff}^{#1}}
\newcommand{\msbar}{$\overline{\mbox{MS}}$}

\newcommand{\gev}{\,\, \mathrm{GeV}}

\newcommand{\re}{\text{Re}\,}
\newcommand{\im}{\text{Im}\,}

\newcommand{\OO}{{\mathcal O}}

\newcommand{\mycaption}[1]{\caption{\sl #1}}

\makeatletter
\def\section{\@startsection {section}{1}{\z@}{-3.5ex plus -1ex minus 
 -.2ex}{2.3ex plus .2ex}{\large\bf\boldmath}}
\def\subsection{\@startsection{subsection}{2}{\z@}{-3.25ex plus -1ex
 minus -.2ex}{1.5ex plus .2ex}{\normalsize\bf\boldmath}}
\def\subsubsection{\@startsection{subsubsection}{3}{\z@}{-3.25ex plus
 -1ex minus -.2ex}{1.5ex plus .2ex}{\normalsize\it}}

\title{GRIFFIN: A C++ library for electroweak radiative corrections in fermion
scattering and decay processes}
\author[1,2]{Lisong Chen}
\author[1]{Ayres Freitas}
\affil[1]{\small Pittsburgh Particle-physics Astro-physics \& Cosmology Center
(PITT-PACC),\\ Department of Physics \& Astronomy, University of Pittsburgh,
Pittsburgh, PA 15260, USA}
\affil[2]{\small Institut f\"{u}r Theoretische Teilchenphysik, Karlsruhe
Institute of Technology (KIT)\\
Wolfgang-Gaede Stra{\ss}e 1, 76128 Karlsruhe, Germany}
\date{} 

\begin{document}

\maketitle

\begin{abstract} 
This paper describes a modular framework for the description of electroweak scattering and decay processes, including but not limited to Z-resonance physics. The framework consistently combines a complex-pole expansion near an s-channel resonance with a regular fixed-order perturbative description away from the resonance in a manifestly gauge-invariant scheme. Leading vertex correction contributions are encapsulated in form factors that can be predicted or treated as numerical fit parameters.

This framework has been implemented in the publicly available object-oriented C++ library GRIFFIN. Version 1.0 of this library provides Standard Model predictions for the IR-subtracted matrix elements for the process $f\bar{f} \to f'\bar{f}'$ with full NNLO and leading higher-order contributions on the Z-resonance, and with NLO corrections off-resonance. The library can straightforwardly be extended to include higher-order corrections, should they become available, or predictions for new physics models. It can be interfaced with Monte-Carlo programs to account for QED and QCD initial-state and final-state radiation.
\end{abstract}


\section{Introduction}

Studies of fermion scattering, $f\bar{f} \to f'\bar{f}'$ for center-of-mass energies near the Z-boson resonance, $\sqrt{s} \sim \MZ$, have played a crucial role in elucidating the structure of the Standard Model (SM) and putting constraints in potential new physics beyond the SM (BSM). These include precision measurements at LEP and SLC (where $f=e$) \cite{ALEPH:2005ab}, as well as Drell-Yan production at the TeVatron and LHC (where $f=u,d,s,c, b$) \cite{CDF:2018cnj,ATLAS:2018gqq,CMS:2018ktx}. Even higher levels of precision can be achieved at the high-luminosity run of the LHC (HL-LHC) \cite{ATLAS:2019mfr} and one of several proposed future $e^+e^-$ colliders: FCC-ee \cite{Abada:2019zxq}, CEPC~\cite{CEPCStudyGroup:2018ghi}, ILC~\cite{Baer:2013cma,Bambade:2019fyw}, CLIC~\cite{Linssen:2012hp,Charles:2018vfv}.

The relevant outcomes of these experiments are typically presented in terms of a set of so-called electroweak pseudo-observables (EWPOs) that encapsulate the dominant radiative corrections in the SM and are most sensitive to BSM physics. Examples of EWPOs are effective Z-fermion couplings, partial Z-boson widths, the effective weak mixing angle $\seff{f}$, and the asymmetry parameters $A_f$; see \emph{e.g.} Ref.~\cite{ALEPH:2005ab} for the definition of these quantities. For the full description of the observable cross-sections, however, a number of other ingredients besides the EWPOs are needed, such as contributions from diagrams without s-channel Z-bosons (\emph{i.e.} s-channel photon exchange and box diagrams) and initial- and final-state QED and QCD radiation effects. A number of software packages provide these ingredients with complete next-to-leading order (NLO) and some partial higher-order corrections included \cite{Bardin:1989tq,Montagna:1998kp,Bardin:1999yd,Arbuzov:2005ma,Bardin:2000kn,Baur:1997wa,Baur:2001ze,Dittmaier:2001ay,Dittmaier:2009cr,Dittmaier:2020vra,Gluza:2004tq,CarloniCalame:2006zq,CarloniCalame:2007cd,Arbuzov:2007db,Bardin:2019zsp,Zykunov:2007zu,Placzek:2013moa,Barze:2013fru}. Among these, the ZFITTER \cite{Bardin:1999yd,Arbuzov:2005ma} and TOPAZ0 \cite{Montagna:1998kp} packages also provide extensive formulae for real photon radiation contributions, including certain selection cuts. They have been widely used in experimental studies. One can refer to \cite{Bertone:2019hks} for the most recent updates on such analytical methods. Alternatively, QED radiation can be simulated with Monte-Carlo (MC) methods. For example, the electroweak corrections provided by the package {\sc Dizet} \cite{Bardin:1989tq}, which is a component of ZFITTER, have been linked to the MC programs KoralZ \cite{Jadach:1999tr} and KKMC \cite{Arbuzov:2020coe}{\footnote{Similar functionality, without using the {\sc Dizet} package, is also provided by a number of other MC tools. See Ref.~\cite{Frixione:2022ofv} for a broader overview.}

However, despite the tremendous success of these software tools, they may not be easily adaptable to future applications that require a higher level of precision. Such applications call for a setup that enables the incorporation of higher-order corrections (NNLO and beyond) in a well-controlled and manifestly gauge-invariant way, as well as a modular object-oriented structure for the straightforward implementation of new SM or BSM contributions. In this article, the new software package GRIFFIN (Gauge-invariant Resonance In Four-Fermion INteractions) is introduced, which aims to provide a framework with these desirable features. It is written in C++ and defines a class hierarchy that can be extended with new results (both in the SM and beyond) without modifying its interface to external users (such as MC or fitting programs). While its object-oriented structure is, in principle, general enough to implement any arbitrary physics process, the current version is focused on 4-fermion processes, \emph{i.e.} scattering processes of the form $f\bar{f} \to f'\bar{f}'$ or decay processes like muon decay. The relevant matrix elements for these processes are implemented in IR-subtracted form, which can be interfaced with the MC program to treat QED and QCD radiation. To describe the Z-boson resonance, it uses a Laurent expansion of the hard matrix elements\footnote{Here ``hard'' refers to the matrix element without initial-state and final-state QED/QCD radiation since the former would produce a deformation of the resonance lineshape.} about the complex pole $s_0 \equiv \MZ^2 - i\MZ\GZ$. Since this pole is an analytical property of the S-matrix, both the pole location and the coefficients of the expansion are individually gauge-invariant \cite{Willenbrock:1991hu,Sirlin:1991fd,Stuart:1991xk,Veltman:1992tm,Gambino:1999ai,Grassi:2001bz}.

The paper is organized as follows. Section~\ref{sec:fprod} introduces the formalism for the complex-pole expansion and discusses what building blocks are required to describe the Z resonance at NNLO precision. On the other hand, outside of the resonance region, no pole expansion is needed.
In section~\ref{sec:fprod2}, it is discussed how on- and off-resonance predictions can be consistently matched to obtain reliable results for $f\bar{f} \to f'\bar{f}'$ at any center-of-mass energy. The implementation of these elements within the GRIFFIN library is described in section~\ref{sec:struct}, and numerical results and comparisons with ZFITTER/{\sc Dizet} are shown in section~\ref{sec:results}. Finally, a summary is provided in section~\ref{sec:summ}.


\section{Fermion pair production on the Z resonance}
\label{sec:fprod}

The matrix element for the process $f\bar{f} \to f'\bar{f}'$ can be decomposed into four different chirality structures, which here will be delineated according to their vector or axial-vector couplings:
\begin{align}
{\cal M} = 
\bigl[ M_{\rm VV} \gamma^\mu \otimes \gamma_\mu
 - M_{\rm VA} \gamma^\mu \otimes \gamma_\mu \gamma^5
 - M_{\rm AV} \gamma^\mu \gamma^5 \otimes \gamma_\mu
 + M_{\rm AA} \gamma^\mu \gamma^5 \otimes \gamma_\mu  \gamma^5 \bigr],
\end{align}
where the $\otimes$ stands for the outer product of two fermion chains. In terms of these quantities, the differential cross-section is given by
\begin{align}
\frac{d\sigma}{d\cos\theta} &=
 \frac{N_c}{32\pi s}|{\cal M}|^2 \\
&= \frac{N_c s}{32\pi}\,
 \Bigl[\begin{aligned}[t]
 &(1+c_\theta^2)
  \bigl(|M_{\rm VV}|^2 + |M_{\rm VA}|^2 + |M_{\rm AV}|^2 + |M_{\rm AA}|^2\bigr) 
  \\[1ex]
 &+ 4c_\theta \, \text{Re}\bigl\{ M_{\rm VV}M_{\rm AA}^* + M_{\rm VA} M_{\rm AV}^*
  \bigr\} \\[.5ex]
 &- 2P_f(1+c_\theta^2)\,\text{Re}\bigl\{ M_{\rm VV}M_{\rm AV}^* + M_{\rm VA}
  M_{\rm AA}^* \bigr\} \\
 &- 4P_fc_\theta \, \text{Re}\bigl\{ M_{\rm VV}M_{\rm VA}^* + M_{\rm AV} 
  M_{\rm AA}^* \bigr\} \Bigr]\,,
 \end{aligned} \label{xsdiff}
\end{align}
where $c_{\theta}=\cos{\theta}$, $\theta$ is the scattering angle in the center-of-mass frame, and $s$ is the center-of-mass energy, $P_f$ is the polarization of the incoming fermion $f$, and the masses of $f$ and $f'$ have been neglected. For Z-boson exchange at tree level, the four chiral matrix elements read
\begin{align}
M_{\rm VV}^{(0)} &= \frac{v^\PZ_{f(0)} v^\PZ_{f'(0)}}{s-s_0}, \quad
M_{\rm VA}^{(0)} = \frac{v^\PZ_{f(0)} a^\PZ_{f'(0)}}{s-s_0}, \quad
M_{\rm AV}^{(0)} = \frac{a^\PZ_{f(0)} v^\PZ_{f'(0)}}{s-s_0}, \quad
M_{\rm AA}^{(0)} = \frac{a^\PZ_{f(0)} a^\PZ_{f'(0)}}{s-s_0}, 
\intertext{where $s_0 \equiv \MZ^2 - i\MZ\GZ$ and}
v^\PZ_{f(0)} &= \frac{eI_f^3(1-4|Q_f|\sw^2)}{2\sw\cw}, \qquad
a^\PZ_{f(0)} = \frac{eI_f^3}{2\sw\cw}
\end{align}
are the vector and axial-vector couplings of the Z-boson to the fermion $f$ ($f = \ell,\nu,u,d,...$). Furthermore, $\sw$ and $\cw$ stand for the sine and cosine of weak-mixing angle.

Note that throughout this document, $\MZ$ and $\GZ$ refer to the mass and width of the Z-boson in the complex-pole scheme, which is theoretically well-defined and gauge-invariant \cite{Willenbrock:1991hu,Sirlin:1991fd,Stuart:1991xk,Veltman:1992tm,Gambino:1999ai,Grassi:2001bz}. However, most experimental measurements are typically reported in terms of the so-called running-width scheme, leading to different values for the mass and width, which we denote as $\mz$ and $\gz$. The two definitions are related via \cite{Bardin:1988xt}\footnote{See Ref.~\cite{Denner:2019vbn} for a general review of the treatment of electroweak gauge-boson resonances.}
\begin{align}
\MZ = \mz\,(1+(\gz/\mz)^2)^{-1/2}, \qquad \GZ = \gz\,[1+(\gz/\mz)^2]^{-1/2}.
\label{massdef}
\end{align}

\bigskip\noindent
In general, when including photon-exchange diagrams and higher-order contributions, the matrix elements can be written as Laurent expansion about the complex pole $s_0$,
\begin{align}
M_{ij} &= \frac{R_{ij}}{s-s_0} + S_{ij} + (s-s_0)S'_{ij} + ...
\qquad (i,j = {\rm V,A}). \label{Mexp}
\end{align}
Note that the scattering angle $\theta$ is kept fixed when expanding $M_{ij}(s,\theta)$.
To construct explicit expressions for $R,S,S'$, we introduce the following quantities:
\begin{align}
Z_{Vf}(s) &\equiv v^\PZ_{f}(s) + v^\gamma_{f}(s)\frac{\Sigma_{\gamma
 \PZ}(s)}{s+\Sigma_{\gamma\gamma}(s)}, &
G_{Vf}(s) &\equiv v^\gamma_{f}(s), \\
Z_{Af}(s) &\equiv a^\PZ_{f}(s) + a^\gamma_{f}(s)\frac{\Sigma_{\gamma
 \PZ}(s)}{s+\Sigma_{\gamma\gamma}(s)}, &
G_{Af}(s) &\equiv a^\gamma_{f}(s),\\
\Sigma_\PZ(s) &\equiv \Sigma_{\PZ\PZ}(s) - \frac{[\Sigma_{\gamma
 Z}(s)]^2}{s+\Sigma_{\gamma\gamma}(s)}\,.
\end{align}
Here $v^{\rm V}_f$ ($a^{\rm V}_f$) is the vector (axial-vector) form factor for the vertex between the gauge boson V (V=Z,$\gamma$) and the fermion $f$, including loop contributions. $\Sigma_{\rm V_1V_2}$ is the self-energy for incoming V$_1$ and outgoing V$_2$ (V$_{1,2}$=Z,$\gamma$). Furthermore, we denote
\begin{align}
B_{ij}(s,t) :&\; \parbox[t]{12cm}{Contribution of $\gamma\gamma$, $ZZ$ and $WW$ box diagrams for initial-state vector/axial-vector current ($i=\rm V,A$) and final-state vector/axial-vector current ($j=\rm V,A$);} \\[1ex]
B_{\gamma\PZ,ij}(s,t) &= \frac{B_{\gamma\PZ,ij}^R}{s-s_0} +
 B_{\gamma\PZ,ij}^S + (s-s_0)B_{\gamma\PZ,ij}^{S'} + ... \;: \notag \\
&\; \parbox[t]{12cm}{Contribution of $\gamma$Z box diagrams, which can also contribute to the leading pole term $R_{ij}$.} \label{gzboxexp}
\end{align}
It should be noted that the coefficients $B_{\gamma\PZ,ij}^{R,S,S',...}$ contain additional logarithms $\ln(1-\frac{s}{s_0})$ that become singular on the pole and need to be accounted for in the Laurent expansion. Up to one-loop order, one thus has $B_{\gamma\PZ,ij}^X = B_{\gamma\PZ,ij}^{X,1} \ln(1-\frac{s}{s_0}) + B_{\gamma\PZ,ij}^{X,0}$ ($X=R,S,S',...$), where $B_{\gamma\PZ,ij}^{X,i}$ are independent of $s$. In practice, these contributions are calculated by expanding the  full analytical expression for the one-loop $\gamma$Z box diagrams while tracking both polynomial and logarithmic singularities.

In terms of the above quantities, the coefficients of the complex-pole expansion are read
\begin{align}
R_{ij} &= \biggl[\frac{Z_{if}Z_{jf'}}{1+\Sigma'_\PZ}\biggr]_{s=s_0} + B^{R}_{\gamma\PZ,ij}\,, \displaybreak[0] \\[1ex]
S_{ij} &= \biggl[ \frac{Z_{if}Z'_{jf'}+Z'_{if}Z_{jf'}}{1+\Sigma'_\PZ}
 - \frac{Z_{if}Z_{jf'}\Sigma''_\PZ}{2(1+\Sigma'_\PZ)^2}
 + \frac{G_{if}G_{jf'}}{s+\Sigma_{\gamma\gamma}}
 + B_{ij} \biggr]_{s=s_0} 
 + B^{S}_{\gamma\PZ,ij}\,, \displaybreak[0] \\[1ex]
S'_{ij} &= \biggl[ 
 \frac{Z_{if}Z''_{jf'}+Z''_{if}Z_{jf'}+2Z'_{if}Z'_{jf'}}{2(1+\Sigma'_\PZ)}
 - \frac{(Z_{if}Z'_{jf'}+Z'_{if}Z_{jf'})\Sigma''_\PZ 
  + \frac{1}{3}Z_{if}Z_{jf'}\Sigma'''_\PZ}{2(1+\Sigma'_\PZ)^2} 
  + \frac{Z_{if}Z_{jf'}(\Sigma''_\PZ)^2}{4(1+\Sigma'_\PZ)^3}\notag \\
&\quad\,\, + \frac{G_{if}G'_{jf'}+G'_{if}G_{jf'}}{s+\Sigma_{\gamma\gamma}}
 - \frac{G_{if}G_{jf'}(1+\Sigma'_{\gamma\gamma})}{(s+\Sigma_{\gamma\gamma})^2}
 + B'_{ij} \biggr]_{s=s_0} 
 + B^{S'}_{\gamma\PZ,ij}\,,
\end{align}
Here $X'$ denotes the derivative of $X$ with respect to $s$.

\bigskip\noindent
The vertex form factors and box diagrams can contain infrared (IR) divergencies from QED and (in the case of external quarks) QCD corrections. When interfacing the matrix elements with a Monte-Carlo (MC) program, these IR divergent contributions and the corresponding real emission contributions will be produced by the MC phase-space generator and showering algorithm. Thus they must be excluded from the hard matrix elements encoded in GRIFFIN.

For the vertex form factors, the IR-divergent contributions can be factorized, $Z_{if}^{\rm tot} = R^i_f \times Z_{if}$, where $R^i_f$ ($i=\rm V,A$) contain the QED/QCD corrections to the $f\bar{f}$ pair (see \emph{e.g.} Ref.~\cite{Chetyrkin:1996ela}).  They are defined via the matrix elements for the decay of a vector boson into $f\bar{f}$:
\begin{align}
    R^{\rm V}_f(s) \equiv \frac{{\cal M}^{\rm QED/QCD}_{V^*\to f\bar{f}}}{{\cal M}^{\rm Born}_{V^*\to f\bar{f}}},
    \qquad
    R^{\rm A}_f(s) \equiv \frac{{\cal M}^{\rm QED/QCD}_{A^*\to f\bar{f}}}{{\cal M}^{\rm Born}_{A^*\to f\bar{f}}}, \label{RVA}
\end{align}
where $V^*$ ($A^*$) denotes a generic vector boson with invariant mass $s$ that couples to the $f\bar{f}$ fermion current with a pure vector (axial-vector) coupling, and the superscript ``QED/QCD'' indicates that all QED and QCD to the desired order are included.
The factorization is not perfect, but the remaining non-factorizable contributions \cite{Czarnecki:1996ei,Harlander:1998cmq} are IR-finite and can be incorporated into $Z_{if}$ order by order. Here and in the following, we adopt the notation that $Z_{if}$ is the IR-finite vertex form factor after the IR-divergent QED/QCD contributions have been factored off, whereas $Z_{if}^{\rm tot}$ is the vertex form factor including all QED/QCD corrections. 

The subtraction of these contributions is less straightforward for the box diagrams, which contain IR-divergent initial-final interference (IFI) terms. We here restrict ourselves to a discussion at NLO, where one encounters IR-divergent IFI terms from two sources, the $\gamma\gamma$ boxes, and the $\gamma$Z boxes. Following the CEEX MC scheme of Ref.~\cite{Jadach:1998jb}, they can be removed with the following subtraction terms:
\begin{align}
&\gamma\gamma \text{ box:}
& B_{\rm VV(1)} &= B^{\rm tot}_{\rm VV(1)} -
 S_{\rm VV}^{(0)}\,\frac{\alpha}{\pi} Q_f Q_{f'} \,
 f_{\rm IR}(m_\gamma,t,u), \\
&\gamma Z \text{ box:}
& B_{\gamma\PZ, ij(1)} &= B^{\rm tot}_{\gamma\PZ, ij(1)}
 - \frac{R_{ij}^{(0)}}{s-s_0}\,\frac{\alpha}{\pi} Q_f Q_{f'} \, [f_{\rm IR}(m_\gamma,t,u)
  + \delta_G(s,t,u)], \label{BgamZsub} \\
&&f_{\rm IR}(m_\gamma,t,u) &= \frac{2\pi}{\alpha}\bigl[R_{e(1)}(t)-R_{e(1)}(u)\bigr]=\ln\Bigl(\frac{1-c_\theta}{1+c_\theta}\Bigr)\biggl[
 \ln\biggl(\frac{2m_\gamma^2}{s\sqrt{1-c^2_\theta}}\biggr) + \frac{1}{2}\biggr],
\notag \displaybreak[0] \\[1ex]
&&\delta_G(s,t,u) &= -2\ln\Bigl(\frac{1-c_\theta}{1+c_\theta}\Bigr)\, \ln\Bigl(\frac{s_0-s}{s_0}\Bigr).
\end{align}
Here the subscripts $(n)$ indicate the loop order. $R_{e(1)}$ is the radiative factor defined in eq.~\eqref{RVA} with one-loop QED corrections (the upper indices for denoting vector/axial vector are suppressed here since the QED correction is chiral-blind). When matching our IR-subtracted results to an MC generator, the $R$ factors can be implemented with any IR regularization scheme in the MC program since they are based on a physical process and thus scheme independent. Only for illustration, we show their form when using a small photon mass $m_\gamma$ as a regulator.} The current version of GRIFFIN uses this subtraction scheme, but other schemes for removing the IR-divergent IFI contributions could also be easily implemented.

\bigskip\noindent
Near the Z resonance, when aiming for a description at N$^n$LO precision, it is typically sufficient to compute only the leading coefficient $R$ to $n$-loop order, whereas $(n-1)$-loop and $(n-2)$-loop precision is adequate for $S$ and $S'$, respectively\footnote{This power counting can be extended to more terms, beyond $S'$, in the Laurent expansion.}.

Furthermore, the ratio $\GZ/\MZ = {\cal O}(\alpha)$, where ${\cal O}(\alpha)$ denotes electroweak NLO corrections, which implies that one can perform expansions in the perturbative order, $\alpha$, and $\GZ/\MZ$ in parallel. For example, $f(s_0) = f(\MZ^2) -i\MZ\GZ\,f'(\MZ^2) -\frac{\MZ^2\GZ^2}{2}\,f''(\MZ^2) + ...$ 
Thus, in summary, we adopt the power counting $(s-s_0)/\MZ^2 \sim \GZ/\MZ \sim \alpha$ for the expansion of the matrix element near the Z pole.

If we wish to expand up to NNLO for the leading pole term, one would in principle, also need the $\gamma$Z box to two-loop order, which is currently unknown. However, it was shown in Refs.~\cite{Melnikov:1995fx,Beenakker:1997bp,Beenakker:1997ir} that at NLO the
total contribution of IFI terms to $R_{ij}$
vanishes when adding up the virtual $\gamma Z$ boxes and real photon radiation
(see also Refs.~\cite{Greco:1980mh,Jadach:1988zp}). This argument, of course, only holds for sufficiently inclusive observables. Furthermore, for quarks in either the initial or final state, Ref.~\cite{Dittmaier:2014qza} demonstrated that the resonance pole also cancels in the IFI contributions for mixed electroweak-QCD NNLO corrections.
A similar argument should apply to the $\gamma\gamma Z$ boxes at electroweak NNLO, although a more careful analysis of this issue would be desirable. Assuming that this argument holds, one only needs to include $B^{R}_{\gamma Z(m)}$, $m=1,...,n-1$ for the computation of $R^{(n)}_{ij}$.

Based on the above considerations, the result for an expansion up to NNLO for the leading pole term $R$ (which implies NLO precision for $S$ and LO for $S'$) reads
\begin{align}
R^{(0)}_{ij} &= Z_{if(0)} Z_{jf'(0)}, \\
R^{(1)}_{ij} &= \bigl[ Z_{if(0)} Z_{jf'(1)} + Z_{if(1)} Z_{jf'(0)}
 - Z_{if(0)} Z_{jf'(0)} \, \Sigma'_{Z(1)} \bigr]_{s=\MZ^2}
 \,, \label{r1} \\
R^{(2)}_{ij} &= \bigl[ Z_{if(0)} Z_{jf'(2)} + Z_{if(2)} Z_{jf'(0)}
 + Z_{if(1)} Z_{jf'(1)} - Z_{if(0)} Z_{jf'(0)} \, \Sigma'_{Z(2)}
 - \Sigma'_{Z(1)} R^{(1)}_{ij} \notag \\
&\quad\; - i\MZ\GZ (Z_{if(0)} Z'_{jf'(1)} + Z'_{if(1)} Z_{jf'(0)}
 - Z_{if(0)} Z_{jf'(0)} \, \Sigma''_{Z(1)}) \bigr]_{s=\MZ^2} 
 + B^R_{\gamma\PZ, ij(1)}
 \,, \displaybreak[0] \\[1ex]
S^{(0)}_{ij} &= \frac{1}{\MZ^2}G_{if(0)} G_{jf'(0)}, \label{coeffS0} \\
S^{(1)}_{ij} &= \biggl[ Z_{if(0)} Z'_{jf'(1)} + Z'_{if(1)} Z_{jf'(0)}
 - \frac{1}{2}Z_{if(0)} Z_{jf'(0)} \, \Sigma''_{Z(1)}
 + \frac{1}{\MZ^2}\bigl(G_{if(0)} G_{jf'(1)} + G_{if(1)} G_{jf'(0)}\bigr) 
 \notag \\
&\quad\; + \frac{i\MZ\GZ -\Sigma_{\gamma\gamma(1)}}{\MZ^4}G_{if(0)} G_{jf'(0)}
 + B_{ij(1)}\biggr]_{s=\MZ^2} + B^{S}_{\gamma\PZ,ij(1)}\,, \label{coeffS1} \displaybreak[0] \\[1ex]
S'^{(0)}_{ij} &= -\frac{1}{\MZ^4}G_{if(0)} G_{jf'(0)}, \label{coeffSp0} 
\end{align}
where the subscripts $(n)$ again indicate the loop order.

\bigskip\noindent
As mentioned in the introduction, electroweak pseudo-observables (EWPOs) are used as an intermediate step when comparing experimental data to theory expectations. The EWPOs can be expressed in terms of the form
factors $F^f_{V,A}$ defined in Ref.~\cite{Freitas:2014hra}\footnote{$F_A^f$ is related to $\rho_f$ introduced in Ref.~\cite{Akhundov:1985fc}, up to a normalization factor.} and in terms of the effective weak mixing
angle $\seff{f}$ (as defined, \emph{e.g.}, in Ref.~\cite{Awramik:2006uz}).
Up to NNLO, and using the power counting $\alpha \sim \GZ/\MZ$, they are given by
\begin{align}
\seff{f} &= \frac{1}{4|Q_f|}\biggl[1-\text{Re}\,\frac{Z_{Vf}}{Z_{Af}}
\biggr]_{s=\MZ^2}, \\
F_A^f &= \biggl[\frac{|Z_{Af}|^2}{1+\text{Re}\,\Sigma'_Z}
 - \frac{1}{2}\MZ\GZ |a^\PZ_{f(0)}|^2 \, \text{Im}\, \Sigma''_Z \biggr]_{s=\MZ^2}
 + {\cal O}(\alpha^3), \\
F_V^f &= \biggl[\frac{|Z_{Vf}|^2}{1+\text{Re}\,\Sigma'_Z}
 - \frac{1}{2}\MZ\GZ |v^\PZ_{f(0)}|^2 \, \text{Im}\, \Sigma''_Z \biggr]_{s=\MZ^2}
 + {\cal O}(\alpha^3) \\
&= F_A^f\biggl[(1-4|Q_f|\seff{f})^2 +
 \Bigl(\text{Im}\,\frac{Z_{Vf}}{Z_{Af}}\Bigr)^2\biggr]
\end{align}
For $f=\nu$ the effective weak mixing angle is ill-defined and irrelevant, and
only $F_A^\nu$ is needed.

The matrix elements for the process $f\bar{f} \to f'\bar{f}'$ can be expressed in terms of these form factors. In fact, they only enter the leading pole coefficient, $R$, as follows:
\begin{align}
R_{ij}^{(0+1+2)} &= 4I^3_fI^3_{f'}\sqrt{F_A^f F_A^{f'}}\,\Bigl[ 
\begin{aligned}[t] &\tilde{Q}_i^f \tilde{Q}_j^{f'} \bigl(1+i\, r^I_{AA} 
 - \tfrac{1}{2}(r^I_{AA})^2 + \tfrac{1}{2}\delta\overline X_{(2)}\bigr) \\
 &+ (\tilde{Q}_i^f I_{j,f'} + \tilde{Q}_j^{f'} I_{i,f})(i - r^I_{AA})
 - I_{i,f}I_{j,f'}\Bigr]
\end{aligned} \notag \\
 &\quad +\MZ\GZ\,Z_{if(0)} Z'_{jf'(0)}\, x^I_{ij}\,, \label{coeffR}
\end{align}
where
\begin{align}
\tilde{Q}^f_V &= 1-4|Q_f|\seff{f}, &
\tilde{Q}^f_A &= 1, \displaybreak[0] \\
I_{V,f} &= \frac{1}{(a^\PZ_{f(0)})^2}\bigl[a^\PZ_{f(0)}\,\text{Im}\, Z_{Vf(1)} 
 - v^\PZ_{f(0)}\, \text{Im}\, Z_{Af(1)} \bigr], 
 & I_{A,f} &= 0, \displaybreak[0] \\
\delta\overline X_{(2)} &= -(\text{Im}\, \Sigma'_{Z(1)})^2
 + 2\,\frac{B^R_{\gamma\PZ, ij(1)}}{R^{(0)}_{ij}}, \displaybreak[0] \\
r^I_{AA} &= \frac{\text{Im}\,Z_{Af(1)}}{a^\PZ_{f(0)}} + 
 \frac{\text{Im}\,Z_{Af'(1)}}{a^\PZ_{f'(0)}} -\text{Im}\,\Sigma'_{Z(1)}, 
  \displaybreak[0] \\
x^I_{ij} &= \frac{\text{Im}\,Z'_{if(1)}}{Z_{if(0)}} + 
 \frac{\text{Im}\,Z'_{jf'(1)}}{Z_{jf'(0)}} -\frac{1}{2}\,\text{Im}\,\Sigma''_{Z(1)}. 
\end{align}


\section{Combination of on- and off-resonance fermion-pair production}
\label{sec:fprod2}

The Laurent series \eqref{Mexp} is only a good approximation in a window of a few GeV about the Z resonance. For values of $\sqrt{s}$ outside of this window, a non-expanded version of the matrix element provides a more accurate description. A unified formulation that works for a wide range of center-of-mass energies near and far from the Z resonance is given by the following prescription:
\begin{align}
M_{ij} &= M_{ij}^{{\rm exp},s_0} + M_{ij}^{\rm noexp} - M_{ij}^{{\rm
exp},\MZ^2}, \label{exps}
\end{align}
where $M_{ij}^{{\rm exp},s_0}$ is the matrix element expanded about the complex
pole $s_0$ as in \eqref{Mexp}, and $M_{ij}^{\rm noexp}$ is the matrix element without any expansion in $s$ and Dyson summation. In other words, it is a straightforward fixed-order matrix element for which the full NLO electroweak corrections are known (see \emph{e.g.} Ref.~\cite{Consoli:1989pc,Boudjema:1996qg}).
To avoid double counting, the
expanded version of the latter, $M_{ij}^{{\rm exp},\MZ^2}$ must be subtracted.
Since $M_{ij}^{\rm noexp}$ has a pole at $s=\MZ^2$, the expansion for
$M_{ij}^{{\rm exp},\MZ^2}$ must be performed about that point\footnote{$M_{ij}^{{\rm exp},s_0}$ contains expansion terms $(s-s_0)^i$ for $i\leq 1$, whereas an expansion of $M_{ij}^{\rm noexp} - M_{ij}^{{\rm exp},\MZ^2}$ would have terms $(s-\MZ^2)^i$ with $i > 1$. Since the two series have different expansion points, the match between them is not perfect, but the mismatch is of order $\OO(\GZ^2/\MZ^2)$ or $\OO(\alpha\GZ/\MZ)$, which is beyond the level of accuracy of our results. \newline We also want to point out that our matching scheme for combining resonant and off-resonant regions is not unique. One could, for instance, use the complex-mass scheme\cite{Denner:2005fg,Denner:2006ic} to calculate the off-resonant matrix elements, but a more careful investigation of this would be needed. At NLO our prescriptions is equivalent to the one in Ref.~\cite{Dittmaier:2009cr}.}:
\begin{align}
M_{ij}^{{\rm exp},\MZ^2} &= \frac{\overline{R}'_{ij}}{(s-\MZ^2)^2} + 
 \frac{\overline{R}_{ij}}{s-\MZ^2} + \overline{S}_{ij} + 
 (s-\MZ^2)\overline{S}'_{ij} + ...
\end{align}
Up to NLO, the coefficients are given by
\begin{align}
\overline{R}'^{(0)}_{ij} &= 0, \qquad
\overline{R}'^{(1)}_{ij} = -Z_{if(0)} Z_{jf'(0)} \, \Sigma_{Z(1)}\big|_{s=\MZ^2}
 \,, \\
\overline{R}^{(0)}_{ij} &= Z_{if(0)} Z_{jf'(0)}, \\
\overline{R}^{(1)}_{ij} &= \bigl[ Z_{if(0)} Z_{jf'(1)} + Z_{if(1)} Z_{jf'(0)}
 - Z_{if(0)} Z_{jf'(0)} \, \Sigma'_{Z(1)} \bigr]_{s=\MZ^2}
 + B^{\overline{R}}_{\gamma\PZ, ij(1)}
 \,, \\
\overline{S}^{(0)}_{ij} &= \frac{1}{\MZ^2}G_{if(0)} G_{jf'(0)}, \\
\overline{S}^{(1)}_{ij} &= \biggl[ Z_{if(0)} Z'_{jf'(1)} + Z'_{if(1)} Z_{jf'(0)}
 - \frac{1}{2}Z_{if(0)} Z_{jf'(0)} \, \Sigma''_{Z(1)}
 + \frac{1}{\MZ^2}\bigl(G_{if(0)} G_{jf'(1)} + G_{if(1)} G_{jf'(0)}\bigr) 
 \notag \\
&\quad\; - \frac{\Sigma_{\gamma\gamma(1)}}{\MZ^4}G_{if(0)} G_{jf'(0)}
 + B_{ij(1)} \biggr]_{s=\MZ^2} + B^{\overline{S}}_{\gamma Z,ij(1)} \,, \displaybreak[0] \\[1ex]
\overline{S}'^{(0)}_{ij} &= -\frac{1}{\MZ^4}G_{if(0)} G_{jf'(0)},
\end{align}
Note the presence of the double-pole term $\overline{R}'$, which is purely
imaginary and does not exist for the complex-pole expansion. Similar to eq.~\eqref{gzboxexp}, the $\gamma$Z box diagrams also contribute to the single-pole term $\overline{R}$,
\begin{align}
B_{\gamma\PZ,ij}(s,t) &= \frac{B_{\gamma\PZ,ij}^{\overline{R}}}{s-\MZ^2} +
 B_{\gamma\PZ,ij}^{\overline{S}} + ...
\end{align}
The difference $M_{ij}^{\rm noexp} - M_{ij}^{{\rm exp},\MZ^2}$ is free of any poles at $s=\MZ^2$ and in fact it vanishes in the limit $s \to \MZ^2$.
All three terms in eq.~\eqref{exps} are separately finite and gauge-invariant.
With currently available results, $M_{ij}^{{\rm exp},s_0}$ can be evaluated to NNLO order near the Z pole, as described in the previous section. The current state of the art for $M_{ij}^{\rm noexp}$ is NLO, so that, for consistency, $M_{ij}^{{\rm exp},\MZ^2}$ should also be computed to NLO.

Both the matrix element coefficients \eqref{coeffR} and \eqref{coeffS0}--\eqref{coeffSp0}, as well as the complete matrix element \eqref{exps} are implemented in the GRIFFIN library.


\section{Structure of the GRIFFIN library}
\label{sec:struct}

The GRIFFIN package provides a framework for a hierarchy of C++ classes to compute in principle, any electroweak observable or pseudo-observable within a given model. The current version implements SM predictions for EWPOs and matrix elements for the process $f\bar{f} \to f'\bar{f}'$, with $f \neq f'$. Still, it is straightforward to include other items as well, including but not limited to:
\begin{itemize}
\item matrix elements for $f\bar{f} \to f\bar{f}$, with the same fermion type in the initial and final state, which includes Bhabha scattering;
\item matrix elements for radiation of additional photons ($f\bar{f} \to f'\bar{f}'\gamma$) and/or fermion pairs ($f\bar{f} \to f'\bar{f}'f''\bar{f}''$), with appropriate subtraction of IR singularities (which is broadly equivalent to the concept of ``electroweak pseudo-parameters'' (EWPP) in section C.3 of Ref.~\cite{Blondel:2018mad});
\item predictions for EWPOs in BSM theories or in terms of effective theory extensions of the SM with higher-dimensional operators;
\end{itemize}
The library contains two base classes:
\begin{itemize}
\item class {\tt inval}, which contains user-provided input parameters for a given model (such as the SM or some extension thereof);
\item class {\tt psobs}, which returns a numerical prediction for an observable or pseudo-observable, for the input parameters provided by an {\tt inval} object.
\end{itemize}
In its basic form, {\tt inval} simply has some basic methods for setting and retrieving the values of some input parameters. However, one can define extended classes derived from {\tt inval} to perform computations of input parameters, such as translating between masses in the complex-pole scheme and the running-width scheme, see eq.~\eqref{massdef}, or computing the W-boson mass from the Fermi constant \cite{Awramik:2003rn}.

\begin{table}[tb]
\renewcommand{\arraystretch}{1.2}
\centering
\begin{tabular}{|cc|ccc|c|}
\hline
\multicolumn{2}{|c|}{Boson masses and widths} & \multicolumn{3}{c|}{Fermion masses} & Couplings \\
\hline
\MW & \GW & $m_{\rm e}^{\rm OS}$ & $m_{\rm d}^{\overline{\text{MS}}}(\MZ)$ &
 $m_{\rm u}^{\overline{\text{MS}}}(\MZ)$ & $\alpha(0)$\\
\MZ & \GZ & $m_{\mu}^{\rm OS}$ & $m_{\rm s}^{\overline{\text{MS}}}(\MZ)$ & $m_{\rm c}^{\overline{\text{MS}}}(\MZ)$ & $\Delta\alpha \equiv 1-\alpha(0)/\alpha(\MZ^2)$ \\
\MH & & $m_{\tau}^{\rm OS}$ & $\mb^{\overline{\text{MS}}}(\MZ)$ & $\mt^{\rm OS}$
 & $\as^{\overline{\text{MS}}}(\MZ)$ \\
&&&&& $G_\mu $ \\
\hline
\end{tabular}
\mycaption{SM input parameters used in GRIFFIN. Here OS and \msbar\ refer to the on-shell and \msbar\ scheme, respectively. For most quantities currently encoded in GRIFFIN, fermion masses besides the top-quark mass are being ignored. The CKM matrix is taken to be the unit matrix. $\alpha(0)$ refers to the electromagnetic coupling in the Thomson limit, and $G_\mu$ is the Fermi constant of muon decay.}
\label{tab:input}
\end{table}
The base version of GRIFFIN defines a set of input parameters for SM calculations, listed in Tab.~\ref{tab:input}. Most of these parameters are defined within the on-shell (OS) renormalization scheme, with the exception of light quark masses and the strong coupling, for which the \msbar\ scheme is assumed (at the scale $\mu=\MZ$). Additional input parameters for flavor physics or BSM scenarios can be easily added.

The user has the option to choose between input classes that either use $\alpha(0),\MW,\MZ$ or $\alpha(0),G_\mu,\MZ$ as inputs to define the electroweak couplings. Here $\alpha(0)$ is the electromagnetic coupling in the Thomson limit, and $G_\mu$ is the Fermi constant of muon decay. An additional input is the shift $\Delta\alpha$ between the running electromagnetic couplings at the scales $q^2=0$ and $q^2=\MZ^2$. $\Delta\alpha$ receives contributions from leptons, which has been computed to four-loop order \cite{Sturm:2013uka}, and from quarks or hadrons, which can be extracted from data \cite{Davier:2019can,Keshavarzi:2019abf,Jegerlehner:2019lxt}.

A child class descending from {\tt psobs} can in principle encode predictions for any observable or pseudo-observable within any given model. The base version of GRIFFIN includes SM predictions for form factors, such as $\seff{f}$ and $F^f_{V,A}$, and for matrix elements for the process $f\bar{f} \to f'\bar{f}'$ near the Z resonance, using the complex pole expansion described in the previous section.

\bigskip\noindent
GRIFFIN version 1.0 contains the following SM corrections:
\begin{itemize}
\item
Complete one-loop corrections for $\seff{f}$ \cite{Marciano:1980pb,Akhundov:1985fc} are implemented in the class {\tt SW\_SMNLO}. On top of this, electroweak \cite{Awramik:2004ge,Hollik:2005va,Awramik:2006ar,Hollik:2006ma,Awramik:2008gi,Dubovyk:2016aqv} and mixed electroweak-QCD \cite{Djouadi:1987gn,Djouadi:1987di,Kniehl:1989yc,Kniehl:1991gu,Djouadi:1993ss} two-loop corrections, as well as partial higher-order corrections are available in the class {\tt SW\_SMNNLO}. The latter include $\OO(\at\as^2)$ \cite{Avdeev:1994db,Chetyrkin:1995ix}, $\OO(\at^2\as)$, $\OO(\at^3)$ 
\cite{vanderBij:2000cg,Faisst:2003px} and $\OO(\at\as^3)$ \cite{Schroder:2005db,Chetyrkin:2006bj,Boughezal:2006xk} corrections in the limit of a large top Yukawa coupling $y_\Pt$, where $\at \equiv y_\Pt^2/(4\pi)$, and leading fermionic three-loop corrections of orders $\OO(\alpha^3)$ and $\OO(\alpha^2\as)$ \cite{Chen:2020xzx,Chen:2020xot}.
In addition, non-factorizable $\OO(\alpha\as)$ $Zq\bar{q}$ vertex contributions \cite{Czarnecki:1996ei,Harlander:1997zb,Fleischer:1992fq,Buchalla:1992zm,Degrassi:1993ij,Chetyrkin:1993jp} are also implemented in {\tt SW\_SMNNLO}.
\item
Similarly, the classes {\tt FA\_SMNLO} and {\tt FV\_SMNLO} provide one-loop corrections \cite{Akhundov:1985fc} for the form factors $F_{V,A}^f$, whereas {\tt FA\_SMNNLO} and {\tt FV\_SMNNLO} contain electroweak \cite{Freitas:2013dpa,Freitas:2014hra,Dubovyk:2018rlg,Dubovyk:2019szj} and mixed electroweak-QCD \cite{Djouadi:1987gn,Djouadi:1987di,Kniehl:1989yc,Kniehl:1991gu,Djouadi:1993ss} two-loop corrections, as well as the partial higher-order corrections and non-factorizable contributions mentioned in the previous bullet point.
\item
For the process $f\bar{f} \to f'\bar{f}'$: The class {\tt mat\_SMNNLO} computes the matrix element according to section~\ref{sec:fprod2} with the following ingredients:
\begin{itemize}
\item
All contributions needed to compute the matrix element coefficient $R$ to NNLO accuracy according to \eqref{coeffR}, and the coefficients $S$ and $S'$ to NLO and LO, respectively, see eqs.~\eqref{coeffS0}--\eqref{coeffSp0}. These are also separately available in the member functions {\tt coeffR, coeffS, coeffSp} of {\tt mat\_SMNNLO}. 
\item
The off-resonance contribution, $M_{ij}^{\rm noexp} - M_{ij}^{{\rm exp},\MZ^2}$, to NLO precision, see section~\ref{sec:fprod2}, which is also separately available via the member function {\tt resoffZ}.
\end{itemize}
\item
When using the input parameter set $\alpha(0),G_\mu,\MZ$, one needs to compute $\MW$ from these quantities according to
\begin{align}
G_\mu = \frac{\pi\alpha}{\sqrt{2}\MW^2(1-\MW^2/\MZ^2)}(1+\Delta r).
\end{align}
Here $\Delta r$ accounts for radiative corrections. The class {\tt dr\_SMNNLO} contains all higher-order corrections discussed in Ref.~\cite{Awramik:2003rn}, plus the leading fermionic three-loop corrections of orders $\OO(\alpha^3)$ and $\OO(\alpha^2\as)$ \cite{Chen:2020xzx,Chen:2020xot}. These corrections are used in the input classes {\tt invalGmu} and {\tt SMvalGmu}.
\end{itemize}
For any of these quantities, QED and QCD corrections on the external legs have been factored out, as explained in detail in section \ref{sec:fprod}. The logic is that QED/QCD effects depend on detector acceptance and selection cuts and are best simulated with MC methods. GRIFFIN could be interfaced with suitable MC tools to provide the hard electroweak matrix elements.


\section{Sample results and comparisons}
\label{sec:results}

In this section, we show numerical comparisons between GRIFFIN and the {\sc Dizet} library of EW radiative corrections \cite{Bardin:1999yd,Arbuzov:2005ma,dizet645} for the EWPOs and the differential cross-section. For the latter, we use some of the computational frameworks of the KKMCee project\footnote{The authors are grateful to S.~Jadach for sharing a suitable test program with us.} \cite{Jadach:2022mbe}.

We first perform a benchmark test of the EWPOs in comparison with {\sc Dizet v 6.45} \cite{dizet645}, including NNLO and leading NNNLO corrections.
In {\sc Dizet}, the form factor is defined as in Eq.~2.4.9 and Eq.~2.4.10 of Ref.~\cite{Bardin:1999yd}.
\begin{equation}
\Gamma_{Z\to f\bar{f}}=\Gamma_0 c_f \big | \rho_Z^f\big |(\big |g_Z^f \big |^2 R_V^f+R_A^f)+\delta_{\alpha\alpha_s}, \label{zfpwdef}
\end{equation}
where we have neglected all lepton masses, $c_f=N_c^f$ is the number of colors, and 
\begin{equation}
\Gamma_0=\frac{G_{\mu} M_Z^3}{24\sqrt{2}\pi}.
\end{equation}
In eq.~\eqref{zfpwdef}, $g_Z^f$ is a complex-valued variables, which in our notation from section~\ref{sec:fprod} is given by
\begin{equation}
g_Z^f=\frac{Z_{Vf}}{Z_{Af}},
\end{equation}
On the other hand, in GRIFFIN, we define the partial width of Z-boson as 
\begin{equation}
\Gamma_{Z\to f\bar{f}}=\frac{N_c^f M_Z}{12 \pi}( F_V^f R_V^f+F_A^f R_A)\,.
\end{equation}
By setting this equal to eq.~\eqref{zfpwdef} we obtain the relation
\begin{equation}
\frac{N_c^f M_Z}{12 \pi}F_A^f\biggl( \frac{F_V^f}{F_A^f}R_V^f +R_A^f\biggr)=\frac{N_c^f M_Z}{12 \pi}F_A^f( \big | g_Z^f \big |^2R_V^f +R_A^f)=\Gamma_0 c_f \big | \rho_Z^f\big |(\big |g_Z^f \big |^2R_V^f +R_A^f),
\end{equation}
This implies the following relation between $|\rho_Z^f|$ and $ F_A^f$:
\begin{equation}
\big | \rho_Z^f\big |= \frac{2\sqrt{2} F_A^f}{G_{\mu}M_Z^2}\label{rhogrif}
\end{equation}
One should notice that the non-factorizable mixed QCD-EW corrections are considered as an additive part to the Z widths in {\sc Dizet}, whereas in {\sc GRIFFIN}, they are absorbed in the form factors $F_{V,A}$. This will cause a small numerical mismatch when comparing the form factor $F_A$ to $\rho_Z^f$. Besides, one also has to notice that eq.~\ref{rhogrif} is the modulus of $\rho_Z^f$ instead of $\re{\rho_Z^f}$. Hence to compare these observables, we need to use both the $\im{\rho_Z^f}$ and $\re{\rho_Z^f}$ output
 from {\sc Dizet} to reconstruct $|\rho_Z^f|$.

The flags used by {\sc Dizet v.6.45} are listed as follows:

\begin{center}
\begin{tabular}{ccccc}
\hline\hline
 \texttt{IHVP}=5 & \texttt{IAMT4}=8 & \texttt{IQCD}=3 & \texttt{IMOMS}=1 & \texttt{IMASS}=0 \\
 \texttt{ISCRE}=0 & \texttt{IALEM}=0 & \texttt{IMASK}=0 & \texttt{ISCAL}=0 & \texttt{IBARB}=2 \\
 \texttt{IFTJR}=1 & \texttt{IFACR}=2 & \texttt{IFACT}=0 & \texttt{IHIGS}=0 & \texttt{IAFMT}=3 \\
 \texttt{IEWLC}=0 & \texttt{ICZAK}=1 & \texttt{IHIGS}=1 & \texttt{IALE2}=3 & \texttt{IGFER}=2 \\
 \texttt{IDDZZ}=1 & \texttt{IAMW2}=1 & \texttt{ISFSR}=1 & \texttt{IDMWW}=0 & \texttt{IDSWW}=0 \\
 \hline\hline
\end{tabular}
\end{center}

\medskip\noindent
Due to the limited options of EW input schemes offered by subroutines in {\sc Dizet}, we have to set $G_{\mu},\, \MZ$ as inputs and use the {\sc Dizet} outputs for $\MW$ and $\Gamma_{\PW,\PZ}$ as inputs for {\sc GRIFFIN}, as shown in the following table:

\bigskip\noindent
\renewcommand{\arraystretch}{1.1}
\begin{tabular}{|l|l|}
\hline
\multicolumn{2}{|c|}{{\sc GRIFFIN} input parameters}  \\ \hline
{\sc Dizet} input parameters & {\sc Dizet} output     \\ \hline
\begin{tabular}[c]{@{}l@{}}$\alpha_s(M_Z^2)=0.118,\quad \alpha=1/137.035999084$\\ $\Delta\alpha=0.059,\quad \MZ=91.1876\gev, \quad G_{\mu}=1.166137\times 10^{-5}$\\ $\mt=173.0\gev, \quad \MH=125.0\gev,\,\quad m_{\rm e,\mu,\tau, u, d, s, c, b}=0\gev$\end{tabular} & \begin{tabular}[c]{@{}l@{}}$\Gamma_Z=2.495890\gev$\\ $M_W=80.3599\gev$\\ $\Gamma_W=2.090095\gev$\end{tabular} \\
\hline
\end{tabular}

\bigskip\noindent
With these inputs, we find the numerical results for the form factors $\Delta r$, $|\rho_Z^f|$ and $\sin^2\theta_{eff}^f$, as well as the partial Z width $\Gamma_{Z\to f\bar{f}}$ shown in Tab.~\ref{tab:zfvsgr}. Both the {\sc Dizet} and {\sc GRIFFIN} results for the form factors include full $\OO({\alpha^2})$ corrections, $\OO({\alpha\alpha_s})$ and $\OO({\alpha\alpha_s^2})$ QCD corrections, and leading higher-order corrections in an expansion in $\mt^2$ of $\OO({\at^3})$, $\OO({\at^2\alpha_s})$, and $\OO({\at\alpha_s^3})$, but with the following differences: (a) the {\sc GRIFFIN} result for $|\rho_Z^f|$ additionally includes non-factorizable EW-QCD corrections; (b) the {\sc Dizet} result for $\Delta r$ does not include the $\OO({\at\as^3})$ contributions, and the $\OO(\alpha\as, \alpha\as^2)$ terms are computed only in a large-$\mt$ approximation. 

A better agreement for $\Delta r$ is obtained when adjusting GRIFFIN to match the order of $\Delta r$ in {\sc Dizet}, by only summing corrections of $\OO(\alpha,\alpha^2,\alpha_t\alpha_s,\alpha_t\alpha_s^2)$. In this case, one finds a 4-digit agreement, as shown in Tab.~\ref{tab:zfvsgr}.

\begin{table}[tb]
\begin{tabular}{|l| lll|}
\hline &&& \\[-2.5ex]
 & {\sc Dizet 6.45} & \parbox{5em}{{\sc GRIFFIN} all orders} & \parbox{8.5em}{{\sc GRIFFIN} $\OO(\alpha,\alpha^2,\alpha_t\alpha_s,\alpha_t\alpha_s^2)$}  \\[1.5ex]
\hline
$\Delta r$ & $3.63947\times 10^{-2}$ & $3.68836\times 10^{-2}$ &
  $3.63987\times 10^{-2}$ \\
\hline 
\end{tabular}

\medskip
\begin{tabular}{| l | c l | c l | c c |}
\hline
               & \multicolumn{2}{c |}{$|{\rho_Z^f}|$} & \multicolumn{2}{c |}{$\seff{f}$} & \multicolumn{2}{c|}{$\Gamma_{Z\to f\bar{f}}$} \\
               \hline
               & {\sc Dizet 6.45}          & {\sc GRIFFIN}          & {\sc Dizet 6.45}               & {\sc GRIFFIN}              & {\sc Dizet 6.45}               & {\sc GRIFFIN}              \\
               \hline
$\nu\bar{\nu}$ & 1.00800         & 1.00814          & 0.231119              & NAN                  & 0.167206            & 0.167197            \\
$\ell \bar{\ell}$   & 1.00510         & 1.00519          & 0.231500           & 0.231534            & 0.083986           & 0.083975           \\
$u \bar{u}$    & 1.00578          & 1.00573          & 0.231393              & 0.231420             & 0.299938            & 0.299958             \\
$d \bar{d}$    & 1.00675         & 1.00651          & 0.231266              & 0.231309             & 0.382877             & 0.382846             \\
$b \bar{b}$    & 0.99692         & 0.99420         & 0.232737              & 0.23292             & 0.376853              & 0.377432          \\
\hline
\end{tabular}
\mycaption{The numerical comparison of the EWPOs and form factors $\rho$ between {\sc Dizet} and {\sc GRIFFIN}. The partial width results are for a single fermion family. See text for details.
}\label{tab:zfvsgr}
\end{table}

Most predictions given by both programs for $|\rho^f_Z|$, $\seff{f}$ and $\Gamma_{\PZ\to f\bar{f}}$ agree with each other by at least four decimal points. As aforementioned, the definition of the effective weak-mixing angle at $f=\nu$ is ill-defined. Owing to an alternative definition of $\seff{f}$ by  {\sc Dizet} (see eq. 5.6 in Ref.~\cite{Arbuzov:2005ma}), a number is yet produced without phenomenological implications.  The discrepancy is mildly larger for the form factors $|\rho^d_Z|$ and $|\rho^b_Z|$ for quark final states, which reflects the different implementations of the non-factorizable EW-QCD corrections (as mentioned above, these are encapsulated in the form factors in {\sc GRIFFIN}, but treated separately in {\sc Dizet}). This is especially true for $|\rho^b_Z|$, where the top quark comes into play for these types of corrections. However, these implementation differences do not affect the predictions for the partial widths at the given order, and indeed one can see from the table the numbers for $\Gamma_{\PZ\to q\bar{q}}$ agree better. 

\begin{figure}[tb]
\includegraphics[width=8cm]{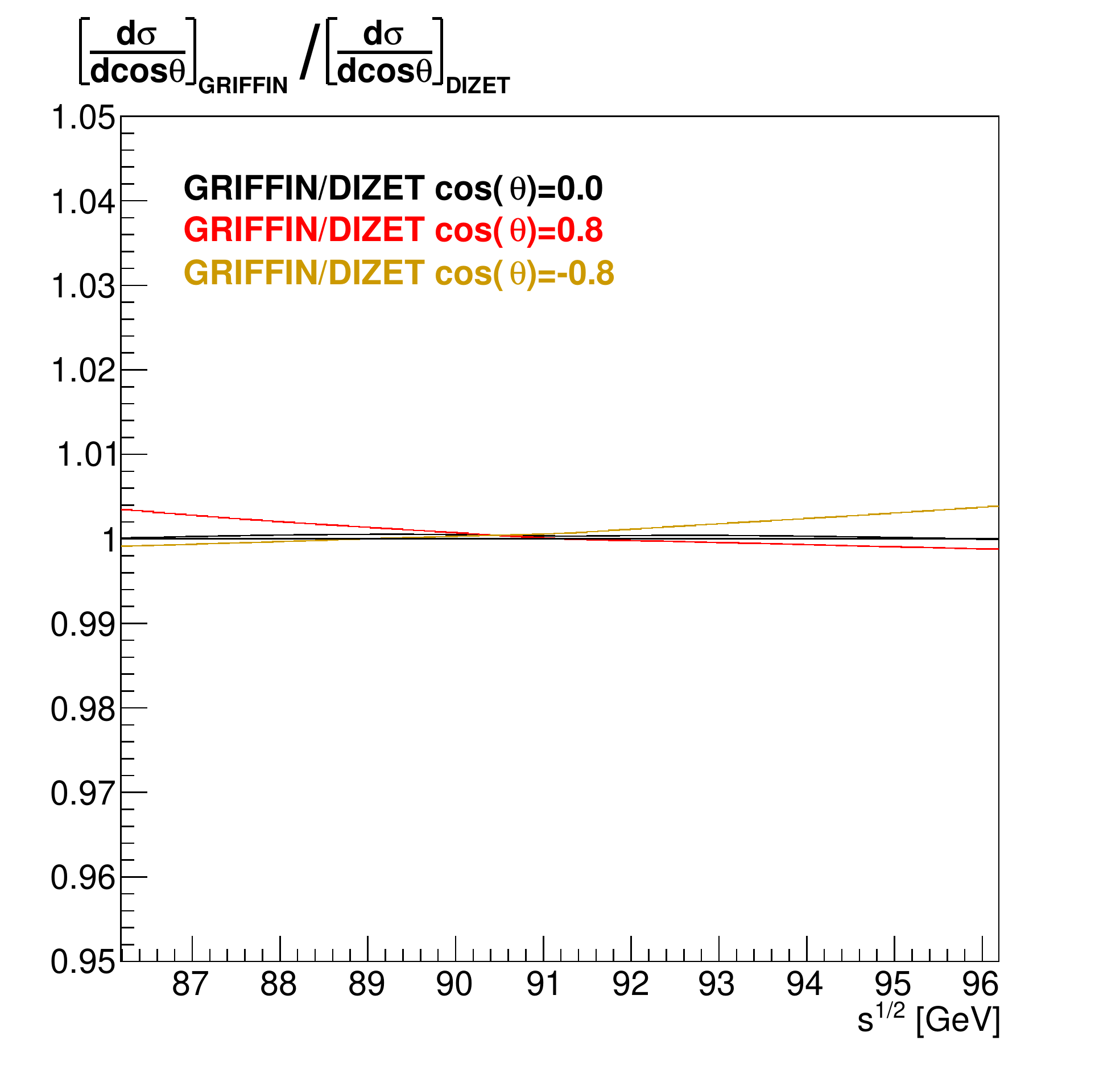}
\includegraphics[width=8cm]{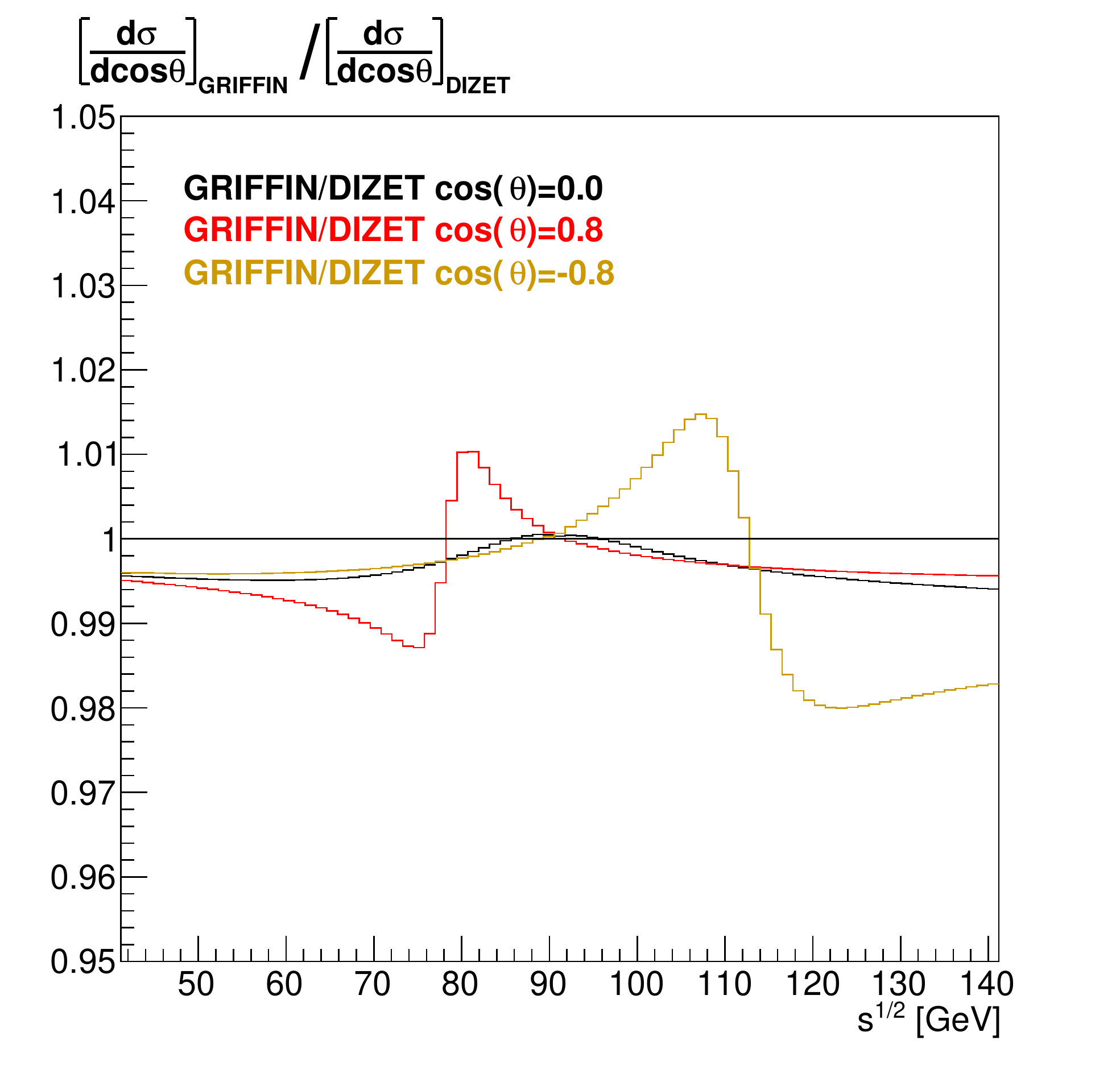}
\mycaption{Ratios of the differential cross-section for $e^+e^- \to \mu^+\mu^-$ using predictions from {\sc GRIFFIN~v1.0} and {\sc Dizet 6.45}, for three choices of the scattering angle $\theta$. The left plot is focused on the Z-pole region, while the right plot shows a wider range of center-of-mass energies.}
\label{fig:dxsec}
\end{figure}

\bigskip\noindent
Let us now move on to comparisons of predictions for the differential cross-section, where for concreteness, we focus on the process $e^+e^- \to \mu^+\mu^-$. Within GRIFFIN, these predictions have been computed using the class {\tt mat\_SMNNLO}, whereas for {\tt Dizet 6.45}, they are based on outputs of the subroutine {\tt ROKANC}, which have been assembled into predictions for the differential cross-section using the KKMCee framework. Given that the $\gamma Z$ box contribution is not included in {\sc Dizet}, we also turned off the $\gamma Z$ contributions in {\sc GRIFFIN} for this comparison for consistency. The results are shown in Fig.~\ref{fig:dxsec}.

The left plot in the figure shows that there is very good agreement between {\sc GRIFFIN~v1.0} and {\sc Dizet 6.45} in the Z-pole region, with deviations of ${\cal O}(10^{-3})$ or less. When analyzing a larger range of center-of-mass energies, as shown in the right plot, one finds larger discrepancies at the level of 0.5--2\%. 
This is not surprising since away from the Z resonances, both codes only deliver NLO precision, and differences in the implementation in {\sc GRIFFIN} and {\sc Dizet} would be of NNLO. In particular, {\sc Dizet} does not use the manifestly gauge-invariant pole expansion scheme described in sections~\ref{sec:fprod} and \ref{sec:fprod2}.
Note that the relative corrections in some kinematic regions (\emph{e.g.} $\cos\theta=+0.8$ below the resonance and $\cos\theta=-0.8$ above the resonance) are enhanced due to strong cancellations between the s-channel photon and Z exchange contributions, which render the tree-level matrix element small. In these regions, the NLO corrections reach 20--30\%, so that ${\cal O}(\%)$ discrepancies from missing NNLO contributions are perfectly consistent with expectations.

We also wish to note that {\sc GRIFFIN} provides a framework that can be systematically extended to higher orders, and NNLO corrections for $e^+e^- \to f\bar{f}$ can be included in the library once they become available. With this improvement, the theory uncertainty for the differential cross-section will likely be reduced significantly below 1\%.


\section{Summary}
\label{sec:summ}

The GRIFFIN library provides a consistent description of the IR-subtracted matrix elements for fermion scattering for a wide range of center-of-mass energies, including close to and far away from the Z-boson resonance. This is achieved by merging a complex-pole expansion, which provides an accurate description near the Z peak, with an unexpanded fixed-order calculation, which is more adequate outside of the Z resonance region.

Version 1.0 of the library includes all currently available higher-order SM corrections for the leading Z-pole term, and NLO SM corrections for the remainder. It also includes two input parameter schemes, which either use the Fermi constant $G_\mu$ or the W mass $\MW$ as inputs and all available higher-order SM corrections for the translation between the two. The results have been validated and compared to the {\sc Dizet 6.45} library.

The structure of GRIFFIN is modular and object-oriented to easily facilitate future extensions. Possible such extensions are:
\begin{itemize}
\item Higher-order corrections for the Z-pole form factors and matrix elements; 
\item Matrix elements for Bhabha scattering and/or final states with additional partons;
\item Predictions for BSM scenarios, including effective theory frameworks such as SMEFT;
\item Implementation of different schemes for factorization of initial- and final-state radiation;
\item Implementation of other processes, such as charged-current Drell-Yan production with a W-boson resonance, or W-boson decays. 
\end{itemize}
The authors invite the community to contact them with feedback, suggestions for improvement, or to contribute new modules to include in the GRIFFIN system.


\section*{Acknowledgments}

The authors are indebted to S.~Jadach and J.~Holeczek for providing us with a test program within the KKMCee framework for producing comparisons, as well as for numerous discussions. We also thank Z.~W\c as for helpful communications. This work has been supported in part by the U.S.~National Science Foundation under grant no.\ PHY-2112829.


\appendix
\section{Appendix: Download information}

The GRIFFIN source code is available at \url{https://github.com/lisongc/GRIFFIN/releases}. The manual is also available at \url{https://github.com/lisongc/GRIFFIN_manual}.


\bibliographystyle{JHEP}{}
\bibliography{griffin}

\providecommand{\href}[2]{#2}\begingroup\raggedright\begin{thebibliography}{10}

\bibitem{ALEPH:2005ab}
{\scshape ALEPH, DELPHI, L3, OPAL, SLD, LEP Electroweak Working Group, SLD
  Electroweak Group, SLD Heavy Flavour Group} collaboration, S.~Schael et~al.,
  \emph{{Precision electroweak measurements on the $Z$ resonance}},
  \href{https://doi.org/10.1016/j.physrep.2005.12.006}{\emph{Phys. Rept.}
  {\bfseries 427} (2006) 257--454},
  [\href{https://arxiv.org/abs/hep-ex/0509008}{{\ttfamily hep-ex/0509008}}].

\bibitem{CDF:2018cnj}
{\scshape CDF, D0} collaboration, T.~A. Aaltonen et~al., \emph{{Tevatron Run II
  combination of the effective leptonic electroweak mixing angle}},
  \href{https://doi.org/10.1103/PhysRevD.97.112007}{\emph{Phys. Rev. D}
  {\bfseries 97} (2018) 112007},
  [\href{https://arxiv.org/abs/1801.06283}{{\ttfamily 1801.06283}}].

\bibitem{ATLAS:2018gqq}
{\scshape ATLAS} collaboration, \emph{{Measurement of the effective leptonic
  weak mixing angle using electron and muon pairs from $Z$-boson decay in the
  ATLAS experiment at $\sqrt s = 8$ TeV}},  ATLAS-CONF-2018-037.

\bibitem{CMS:2018ktx}
{\scshape CMS} collaboration, A.~M. Sirunyan et~al., \emph{{Measurement of the
  weak mixing angle using the forward-backward asymmetry of Drell-Yan events in
  pp collisions at 8 TeV}},
  \href{https://doi.org/10.1140/epjc/s10052-018-6148-7}{\emph{Eur. Phys. J. C}
  {\bfseries 78} (2018) 701},
  [\href{https://arxiv.org/abs/1806.00863}{{\ttfamily 1806.00863}}].

\bibitem{ATLAS:2019mfr}
{\scshape ATLAS, CMS} collaboration, \emph{{Addendum to the report on the
  physics at the HL-LHC, and perspectives for the HE-LHC: Collection of notes
  from ATLAS and CMS}},
  \href{https://doi.org/10.23731/CYRM-2019-007.Addendum}{\emph{CERN Yellow Rep.
  Monogr.} {\bfseries 7} (2019) Addendum},
  [\href{https://arxiv.org/abs/1902.10229}{{\ttfamily 1902.10229}}].

\bibitem{Abada:2019zxq}
{\scshape FCC} collaboration, A.~Abada et~al., \emph{{FCC-ee: The Lepton
  Collider}: {Future Circular Collider Conceptual Design Report Volume 2}},
  \href{https://doi.org/10.1140/epjst/e2019-900045-4}{\emph{Eur. Phys. J. ST}
  {\bfseries 228} (2019) 261--623}.

\bibitem{CEPCStudyGroup:2018ghi}
{\scshape CEPC Study Group} collaboration, {CEPC Study Group, J. Guimarães da
  Costa, Y. Gao, S. Jin, J. Qian, C.~Tully, C.~Young, L.~Wang, M. Ruan, H. Zhu,
  Q. Ouyang et al.}, \emph{{CEPC Conceptual Design Report: Volume 2 - Physics
  \& Detector}},  \href{https://arxiv.org/abs/1811.10545}{{\ttfamily
  1811.10545}}.

\bibitem{Baer:2013cma}
H.~Baer, T.~Barklow, K.~Fujii, Y.~Gao, A.~Hoang, S.~Kanemura et~al., \emph{{The
  International Linear Collider Technical Design Report - Volume 2: Physics,
  }},  \href{https://arxiv.org/abs/1306.6352}{{\ttfamily 1306.6352}}.

\bibitem{Bambade:2019fyw}
P.~Bambade et~al., \emph{{The International Linear Collider: A Global Project,
  }},  \href{https://arxiv.org/abs/1903.01629}{{\ttfamily 1903.01629}}.

\bibitem{Linssen:2012hp}
\emph{{Physics and Detectors at CLIC: CLIC Conceptual Design Report}},
  \href{https://arxiv.org/abs/1202.5940}{{\ttfamily 1202.5940}}.

\bibitem{Charles:2018vfv}
{\scshape CLICdp, CLIC} collaboration, T.~Charles et~al., \emph{{The Compact
  Linear Collider (CLIC) - 2018 Summary Report}},
  \href{https://arxiv.org/abs/1812.06018}{{\ttfamily 1812.06018}}.

\bibitem{Bardin:1989tq}
D.~Y. Bardin, M.~S. Bilenky, T.~Riemann, M.~Sachwitz and H.~Vogt, \emph{{Dizet:
  A Program Package for the Calculation of Electroweak One Loop Corrections for
  the Process e+ e- ---\ensuremath{>} f+ f- Around the Z0 Peak}},
  \href{https://doi.org/10.1016/0010-4655(90)90179-5}{\emph{Comput. Phys.
  Commun.} {\bfseries 59} (1990) 303--312}.

\bibitem{Montagna:1998kp}
G.~Montagna, O.~Nicrosini, F.~Piccinini and G.~Passarino, \emph{{TOPAZ0 4.0: A
  New version of a computer program for evaluation of deconvoluted and
  realistic observables at LEP-1 and LEP-2}},
  \href{https://doi.org/10.1016/S0010-4655(98)00080-0}{\emph{Comput. Phys.
  Commun.} {\bfseries 117} (1999) 278--289},
  [\href{https://arxiv.org/abs/hep-ph/9804211}{{\ttfamily hep-ph/9804211}}].

\bibitem{Bardin:1999yd}
D.~Y. Bardin, P.~Christova, M.~Jack, L.~Kalinovskaya, A.~Olchevski, S.~Riemann
  et~al., \emph{{ZFITTER v.6.21: A Semianalytical program for fermion pair
  production in $e^+ e^-$ annihilation}},
  \href{https://doi.org/10.1016/S0010-4655(00)00152-1}{\emph{Comput. Phys.
  Commun.} {\bfseries 133} (2001) 229--395},
  [\href{https://arxiv.org/abs/hep-ph/9908433}{{\ttfamily hep-ph/9908433}}].

\bibitem{Arbuzov:2005ma}
A.~B. Arbuzov, M.~Awramik, M.~Czakon, A.~Freitas, M.~W. Grunewald, K.~Monig
  et~al., \emph{{ZFITTER: A Semi-analytical program for fermion pair production
  in e+ e- annihilation, from version 6.21 to version 6.42}},
  \href{https://doi.org/10.1016/j.cpc.2005.12.009}{\emph{Comput. Phys. Commun.}
  {\bfseries 174} (2006) 728--758},
  [\href{https://arxiv.org/abs/hep-ph/0507146}{{\ttfamily hep-ph/0507146}}].

\bibitem{Bardin:2000kn}
D.~Y. Bardin, L.~Kalinovskaya and G.~Nanava, \emph{{An Electroweak library for
  the calculation of EWRC to e+ e- ---\ensuremath{>} f anti-f within the topfit
  project}},  \href{https://arxiv.org/abs/hep-ph/0012080}{{\ttfamily
  hep-ph/0012080}}.

\bibitem{Baur:1997wa}
U.~Baur, S.~Keller and W.~K. Sakumoto, \emph{{QED radiative corrections to $Z$
  boson production and the forward backward asymmetry at hadron colliders}},
  \href{https://doi.org/10.1103/PhysRevD.57.199}{\emph{Phys. Rev. D} {\bfseries
  57} (1998) 199--215}, [\href{https://arxiv.org/abs/hep-ph/9707301}{{\ttfamily
  hep-ph/9707301}}].

\bibitem{Baur:2001ze}
U.~Baur, O.~Brein, W.~Hollik, C.~Schappacher and D.~Wackeroth,
  \emph{{Electroweak radiative corrections to neutral current Drell-Yan
  processes at hadron colliders}},
  \href{https://doi.org/10.1103/PhysRevD.65.033007}{\emph{Phys. Rev. D}
  {\bfseries 65} (2002) 033007},
  [\href{https://arxiv.org/abs/hep-ph/0108274}{{\ttfamily hep-ph/0108274}}].

\bibitem{Dittmaier:2001ay}
S.~Dittmaier and M.~Kr\"amer, \emph{{Electroweak radiative corrections to W
  boson production at hadron colliders}},
  \href{https://doi.org/10.1103/PhysRevD.65.073007}{\emph{Phys. Rev. D}
  {\bfseries 65} (2002) 073007},
  [\href{https://arxiv.org/abs/hep-ph/0109062}{{\ttfamily hep-ph/0109062}}].

\bibitem{Dittmaier:2009cr}
S.~Dittmaier and M.~Huber, \emph{{Radiative corrections to the neutral-current
  Drell-Yan process in the Standard Model and its minimal supersymmetric
  extension}}, \href{https://doi.org/10.1007/JHEP01(2010)060}{\emph{JHEP}
  {\bfseries 01} (2010) 060},
  [\href{https://arxiv.org/abs/0911.2329}{{\ttfamily 0911.2329}}].

\bibitem{Dittmaier:2020vra}
S.~Dittmaier, T.~Schmidt and J.~Schwarz, \emph{{Mixed NNLO
  QCD\texttimes{}electroweak corrections of $\mathcal{O}(N_f \alpha_s \alpha)$
  to single-W/Z production at the LHC}},
  \href{https://doi.org/10.1007/JHEP12(2020)201}{\emph{JHEP} {\bfseries 12}
  (2020) 201}, [\href{https://arxiv.org/abs/2009.02229}{{\ttfamily
  2009.02229}}].

\bibitem{Gluza:2004tq}
J.~Gluza, A.~Lorca and T.~Riemann, \emph{{Automated use of DIANA for
  two-fermion production at colliders}},
  \href{https://doi.org/10.1016/j.nima.2004.07.103}{\emph{Nucl. Instrum. Meth.
  A} {\bfseries 534} (2004) 289--292},
  [\href{https://arxiv.org/abs/hep-ph/0409011}{{\ttfamily hep-ph/0409011}}].

\bibitem{CarloniCalame:2006zq}
C.~M. Carloni~Calame, G.~Montagna, O.~Nicrosini and A.~Vicini, \emph{{Precision
  electroweak calculation of the charged current Drell-Yan process}},
  \href{https://doi.org/10.1088/1126-6708/2006/12/016}{\emph{JHEP} {\bfseries
  12} (2006) 016}, [\href{https://arxiv.org/abs/hep-ph/0609170}{{\ttfamily
  hep-ph/0609170}}].

\bibitem{CarloniCalame:2007cd}
C.~M. Carloni~Calame, G.~Montagna, O.~Nicrosini and A.~Vicini, \emph{{Precision
  electroweak calculation of the production of a high transverse-momentum
  lepton pair at hadron colliders}},
  \href{https://doi.org/10.1088/1126-6708/2007/10/109}{\emph{JHEP} {\bfseries
  10} (2007) 109}, [\href{https://arxiv.org/abs/0710.1722}{{\ttfamily
  0710.1722}}].

\bibitem{Arbuzov:2007db}
A.~Arbuzov, D.~Bardin, S.~Bondarenko, P.~Christova, L.~Kalinovskaya, G.~Nanava
  et~al., \emph{{One-loop corrections to the Drell--Yan process in SANC. (II).
  The Neutral current case}},
  \href{https://doi.org/10.1140/epjc/s10052-008-0531-8}{\emph{Eur. Phys. J. C}
  {\bfseries 54} (2008) 451--460},
  [\href{https://arxiv.org/abs/0711.0625}{{\ttfamily 0711.0625}}].

\bibitem{Bardin:2019zsp}
D.~Y. Bardin et~al., \emph{{Precision Description of Processes at Colliders in
  the SANC System}},
  \href{https://doi.org/10.1134/S1063779619040026}{\emph{Phys. Part. Nucl.}
  {\bfseries 50} (2019) 395--432}.

\bibitem{Zykunov:2007zu}
V.~A. Zykunov, \emph{{Electroweak corrections to the Drell-Yan process in the
  high dimuon mass range}},
  \href{https://arxiv.org/abs/hep-ph/0702203}{{\ttfamily hep-ph/0702203}}.

\bibitem{Placzek:2013moa}
W.~P\l{}aczek, S.~Jadach and M.~W. Krasny, \emph{{Drell-Yan processes with
  WINHAC}}, \href{https://doi.org/10.5506/APhysPolB.44.2171}{\emph{Acta Phys.
  Polon. B} {\bfseries 44} (2013) 2171--2178},
  [\href{https://arxiv.org/abs/1310.5994}{{\ttfamily 1310.5994}}].

\bibitem{Barze:2013fru}
L.~Barze, G.~Montagna, P.~Nason, O.~Nicrosini, F.~Piccinini and A.~Vicini,
  \emph{{Neutral current Drell-Yan with combined QCD and electroweak
  corrections in the POWHEG BOX}},
  \href{https://doi.org/10.1140/epjc/s10052-013-2474-y}{\emph{Eur. Phys. J. C}
  {\bfseries 73} (2013) 2474},
  [\href{https://arxiv.org/abs/1302.4606}{{\ttfamily 1302.4606}}].

\bibitem{Bertone:2019hks}
V.~Bertone, M.~Cacciari, S.~Frixione and G.~Stagnitto, \emph{{The partonic
  structure of the electron at the next-to-leading logarithmic accuracy in
  QED}}, \href{https://doi.org/10.1007/JHEP03(2020)135}{\emph{JHEP} {\bfseries
  03} (2020) 135}, [\href{https://arxiv.org/abs/1911.12040}{{\ttfamily
  1911.12040}}].

\bibitem{Jadach:1999tr}
S.~Jadach, B.~F.~L. Ward and Z.~Was, \emph{{The Monte Carlo program KORALZ, for
  the lepton or quark pair production at LEP / SLC energies: From version 4.0
  to version 4.04}},
  \href{https://doi.org/10.1016/S0010-4655(99)00437-3}{\emph{Comput. Phys.
  Commun.} {\bfseries 124} (2000) 233--237},
  [\href{https://arxiv.org/abs/hep-ph/9905205}{{\ttfamily hep-ph/9905205}}].

\bibitem{Arbuzov:2020coe}
A.~Arbuzov, S.~Jadach, Z.~W\k{a}s, B.~F.~L. Ward and S.~A. Yost, \emph{{The
  Monte Carlo Program KKMC , for the Lepton or Quark Pair Production at LEP/SLC
  Energies\textemdash{}Updates of electroweak calculations}},
  \href{https://doi.org/10.1016/j.cpc.2020.107734}{\emph{Comput. Phys. Commun.}
  {\bfseries 260} (2021) 107734},
  [\href{https://arxiv.org/abs/2007.07964}{{\ttfamily 2007.07964}}].

\bibitem{Frixione:2022ofv}
S.~Frixione et~al., \emph{{Initial state QED radiation aspects for future
  $e^+e^-$ colliders}},  in \emph{{2022 Snowmass Summer Study}}, 3, 2022,
  \href{https://arxiv.org/abs/2203.12557}{{\ttfamily 2203.12557}}.

\bibitem{Willenbrock:1991hu}
S.~Willenbrock and G.~Valencia, \emph{{On the definition of the Z boson mass}},
  \href{https://doi.org/10.1016/0370-2693(91)90843-F}{\emph{Phys. Lett. B}
  {\bfseries 259} (1991) 373--376}.

\bibitem{Sirlin:1991fd}
A.~Sirlin, \emph{{Theoretical considerations concerning the Z0 mass}},
  \href{https://doi.org/10.1103/PhysRevLett.67.2127}{\emph{Phys. Rev. Lett.}
  {\bfseries 67} (1991) 2127--2130}.

\bibitem{Stuart:1991xk}
R.~G. Stuart, \emph{{Gauge invariance, analyticity and physical observables at
  the Z0 resonance}},
  \href{https://doi.org/10.1016/0370-2693(91)90653-8}{\emph{Phys. Lett. B}
  {\bfseries 262} (1991) 113--119}.

\bibitem{Veltman:1992tm}
H.~G.~J. Veltman, \emph{{Mass and width of unstable gauge bosons}},
  \href{https://doi.org/10.1007/BF01559523}{\emph{Z. Phys. C} {\bfseries 62}
  (1994) 35--52}.

\bibitem{Gambino:1999ai}
P.~Gambino and P.~A. Grassi, \emph{{The Nielsen identities of the SM and the
  definition of mass}},
  \href{https://doi.org/10.1103/PhysRevD.62.076002}{\emph{Phys. Rev. D}
  {\bfseries 62} (2000) 076002},
  [\href{https://arxiv.org/abs/hep-ph/9907254}{{\ttfamily hep-ph/9907254}}].

\bibitem{Grassi:2001bz}
P.~A. Grassi, B.~A. Kniehl and A.~Sirlin, \emph{{Width and partial widths of
  unstable particles in the light of the Nielsen identities}},
  \href{https://doi.org/10.1103/PhysRevD.65.085001}{\emph{Phys. Rev. D}
  {\bfseries 65} (2002) 085001},
  [\href{https://arxiv.org/abs/hep-ph/0109228}{{\ttfamily hep-ph/0109228}}].

\bibitem{Bardin:1988xt}
D.~Y. Bardin, A.~Leike, T.~Riemann and M.~Sachwitz, \emph{{Energy Dependent
  Width Effects in e+ e- Annihilation Near the Z Boson Pole}},
  \href{https://doi.org/10.1016/0370-2693(88)91627-9}{\emph{Phys. Lett. B}
  {\bfseries 206} (1988) 539--542}.

\bibitem{Denner:2019vbn}
A.~Denner and S.~Dittmaier, \emph{{Electroweak Radiative Corrections for
  Collider Physics}},
  \href{https://doi.org/10.1016/j.physrep.2020.04.001}{\emph{Phys. Rept.}
  {\bfseries 864} (2020) 1--163},
  [\href{https://arxiv.org/abs/1912.06823}{{\ttfamily 1912.06823}}].

\bibitem{Chetyrkin:1996ela}
K.~G. Chetyrkin, J.~H. Kuhn and A.~Kwiatkowski, \emph{{QCD corrections to the
  $e^{+} e^{-}$ cross-section and the $Z$ boson decay rate}},
  \href{https://doi.org/10.1016/S0370-1573(96)00012-9}{\emph{Phys. Rept.}
  {\bfseries 277} (1996) 189--281},
  [\href{https://arxiv.org/abs/hep-ph/9503396}{{\ttfamily hep-ph/9503396}}].

\bibitem{Czarnecki:1996ei}
A.~Czarnecki and J.~H. Kuhn, \emph{{Nonfactorizable QCD and electroweak
  corrections to the hadronic Z boson decay rate}},
  \href{https://doi.org/10.1103/PhysRevLett.77.3955}{\emph{Phys. Rev. Lett.}
  {\bfseries 77} (1996) 3955--3958},
  [\href{https://arxiv.org/abs/hep-ph/9608366}{{\ttfamily hep-ph/9608366}}].

\bibitem{Harlander:1998cmq}
R.~Harlander, T.~Seidensticker and M.~Steinhauser, \emph{{Complete corrections
  of Order alpha alpha-s to the decay of the Z boson into bottom quarks}},
  \href{https://doi.org/10.1016/S0370-2693(98)00220-2}{\emph{Phys. Lett. B}
  {\bfseries 426} (1998) 125--132},
  [\href{https://arxiv.org/abs/hep-ph/9712228}{{\ttfamily hep-ph/9712228}}].

\bibitem{Jadach:1998jb}
S.~Jadach, B.~F.~L. Ward and Z.~Was, \emph{{Coherent exclusive exponentiation
  CEEX: The Case of the resonant e+ e- collision}},
  \href{https://doi.org/10.1016/S0370-2693(99)00038-6}{\emph{Phys. Lett. B}
  {\bfseries 449} (1999) 97--108},
  [\href{https://arxiv.org/abs/hep-ph/9905453}{{\ttfamily hep-ph/9905453}}].

\bibitem{Melnikov:1995fx}
K.~Melnikov and O.~I. Yakovlev, \emph{{Final state interaction in the
  production of heavy unstable particles}},
  \href{https://doi.org/10.1016/0550-3213(96)00151-4}{\emph{Nucl. Phys. B}
  {\bfseries 471} (1996) 90--120},
  [\href{https://arxiv.org/abs/hep-ph/9501358}{{\ttfamily hep-ph/9501358}}].

\bibitem{Beenakker:1997bp}
W.~Beenakker, A.~P. Chapovsky and F.~A. Berends, \emph{{Nonfactorizable
  corrections to W pair production}},
  \href{https://doi.org/10.1016/S0370-2693(97)01010-1}{\emph{Phys. Lett. B}
  {\bfseries 411} (1997) 203--210},
  [\href{https://arxiv.org/abs/hep-ph/9706339}{{\ttfamily hep-ph/9706339}}].

\bibitem{Beenakker:1997ir}
W.~Beenakker, A.~P. Chapovsky and F.~A. Berends, \emph{{Nonfactorizable
  corrections to W pair production: Methods and analytic results}},
  \href{https://doi.org/10.1016/S0550-3213(97)00628-7}{\emph{Nucl. Phys. B}
  {\bfseries 508} (1997) 17--63},
  [\href{https://arxiv.org/abs/hep-ph/9707326}{{\ttfamily hep-ph/9707326}}].

\bibitem{Greco:1980mh}
M.~Greco, G.~Pancheri-Srivastava and Y.~Srivastava, \emph{{Radiative
  Corrections to e+ e- ---\ensuremath{>} mu+ mu- Around the Z0}},
  \href{https://doi.org/10.1016/0550-3213(82)90458-8}{\emph{Nucl. Phys. B}
  {\bfseries 171} (1980) 118}.

\bibitem{Jadach:1988zp}
S.~Jadach and Z.~Was, \emph{{Suppression of {QED} Interference Contributions to
  the Charge Asymmetry at the $Z^0$ Resonance}},
  \href{https://doi.org/10.1016/0370-2693(89)90847-2}{\emph{Phys. Lett. B}
  {\bfseries 219} (1989) 103--106}.

\bibitem{Dittmaier:2014qza}
S.~Dittmaier, A.~Huss and C.~Schwinn, \emph{{Mixed QCD-electroweak
  $\mathcal{O}(\alpha_s\alpha)$ corrections to Drell-Yan processes in the
  resonance region: pole approximation and non-factorizable corrections}},
  \href{https://doi.org/10.1016/j.nuclphysb.2014.05.027}{\emph{Nucl. Phys. B}
  {\bfseries 885} (2014) 318--372},
  [\href{https://arxiv.org/abs/1403.3216}{{\ttfamily 1403.3216}}].

\bibitem{Freitas:2014hra}
A.~Freitas, \emph{{Higher-order electroweak corrections to the partial widths
  and branching ratios of the Z boson}},
  \href{https://doi.org/10.1007/JHEP04(2014)070}{\emph{JHEP} {\bfseries 04}
  (2014) 070}, [\href{https://arxiv.org/abs/1401.2447}{{\ttfamily 1401.2447}}].

\bibitem{Akhundov:1985fc}
A.~A. Akhundov, D.~Y. Bardin and T.~Riemann, \emph{{Electroweak One Loop
  Corrections to the Decay of the Neutral Vector Boson}},
  \href{https://doi.org/10.1016/0550-3213(86)90014-3}{\emph{Nucl. Phys. B}
  {\bfseries 276} (1986) 1--13}.

\bibitem{Awramik:2006uz}
M.~Awramik, M.~Czakon and A.~Freitas, \emph{{Electroweak two-loop corrections
  to the effective weak mixing angle}},
  \href{https://doi.org/10.1088/1126-6708/2006/11/048}{\emph{JHEP} {\bfseries
  11} (2006) 048}, [\href{https://arxiv.org/abs/hep-ph/0608099}{{\ttfamily
  hep-ph/0608099}}].

\bibitem{Consoli:1989pc}
M.~Consoli, W.~Hollik and F.~Jegerlehner, \emph{{Electroweak Radiative
  Corrections for $Z$ Physics}},  in \emph{{LEP Physics Workshop}},
  CERN-TH-5527-89, 1989.

\bibitem{Boudjema:1996qg}
F.~Boudjema et~al., \emph{{Standard model processes}},  in \emph{{AGS / RHIC
  Users Annual Meeting}}, 1, 1996,
  \href{https://arxiv.org/abs/hep-ph/9601224}{{\ttfamily hep-ph/9601224}}.

\bibitem{Denner:2005fg}
A.~Denner, S.~Dittmaier, M.~Roth and L.~H. Wieders, \emph{{Electroweak
  corrections to charged-current e+ e- ---\ensuremath{>} 4 fermion processes:
  Technical details and further results}},
  \href{https://doi.org/10.1016/j.nuclphysb.2011.09.001}{\emph{Nucl. Phys. B}
  {\bfseries 724} (2005) 247--294},
  [\href{https://arxiv.org/abs/hep-ph/0505042}{{\ttfamily hep-ph/0505042}}].

\bibitem{Denner:2006ic}
A.~Denner and S.~Dittmaier, \emph{{The Complex-mass scheme for perturbative
  calculations with unstable particles}},
  \href{https://doi.org/10.1016/j.nuclphysbps.2006.09.025}{\emph{Nucl. Phys. B
  Proc. Suppl.} {\bfseries 160} (2006) 22--26},
  [\href{https://arxiv.org/abs/hep-ph/0605312}{{\ttfamily hep-ph/0605312}}].

\bibitem{Blondel:2018mad}
A.~Blondel et~al., \emph{{Standard model theory for the FCC-ee Tera-Z stage}},
  in \emph{{Mini Workshop on Precision EW and QCD Calculations for the FCC
  Studies : Methods and Techniques}}, vol.~3/2019 of \emph{CERN Yellow Reports:
  Monographs}, (Geneva), CERN, 9, 2018,
  \href{https://arxiv.org/abs/1809.01830}{{\ttfamily 1809.01830}},
  \href{https://doi.org/10.23731/CYRM-2019-003}{DOI}.

\bibitem{Awramik:2003rn}
M.~Awramik, M.~Czakon, A.~Freitas and G.~Weiglein, \emph{{Precise prediction
  for the W boson mass in the standard model}},
  \href{https://doi.org/10.1103/PhysRevD.69.053006}{\emph{Phys. Rev. D}
  {\bfseries 69} (2004) 053006},
  [\href{https://arxiv.org/abs/hep-ph/0311148}{{\ttfamily hep-ph/0311148}}].

\bibitem{Sturm:2013uka}
C.~Sturm, \emph{{Leptonic contributions to the effective electromagnetic
  coupling at four-loop order in QED}},
  \href{https://doi.org/10.1016/j.nuclphysb.2013.06.009}{\emph{Nucl. Phys. B}
  {\bfseries 874} (2013) 698--719},
  [\href{https://arxiv.org/abs/1305.0581}{{\ttfamily 1305.0581}}].

\bibitem{Davier:2019can}
M.~Davier, A.~Hoecker, B.~Malaescu and Z.~Zhang, \emph{{A new evaluation of the
  hadronic vacuum polarisation contributions to the muon anomalous magnetic
  moment and to $\mathbf{\boldsymbol\alpha(m_Z^2)}$}},
  \href{https://doi.org/10.1140/epjc/s10052-020-7792-2}{\emph{Eur. Phys. J. C}
  {\bfseries 80} (2020) 241},
  [\href{https://arxiv.org/abs/1908.00921}{{\ttfamily 1908.00921}}].

\bibitem{Keshavarzi:2019abf}
A.~Keshavarzi, D.~Nomura and T.~Teubner, \emph{{$g-2$ of charged leptons,
  $\alpha (M^2_Z)$ , and the hyperfine splitting of muonium}},
  \href{https://doi.org/10.1103/PhysRevD.101.014029}{\emph{Phys. Rev. D}
  {\bfseries 101} (2020) 014029},
  [\href{https://arxiv.org/abs/1911.00367}{{\ttfamily 1911.00367}}].

\bibitem{Jegerlehner:2019lxt}
F.~Jegerlehner, \emph{{$\alpha_{QED, eff}$(s) for precision physics at the
  FCC-ee/ILC}},  in \emph{{Theory for the FCC-ee}: {Report on the 11th FCC-ee
  Workshop Theory and Experiments}} (A.~Blondel, J.~Gluza, S.~Jadach, P.~Janot
  and T.~Riemann, eds.), vol.~3/2020 of \emph{CERN Yellow Reports: Monographs},
  pp.~9--37, 2020, \href{https://doi.org/10.23731/CYRM-2020-003.9}{DOI}.

\bibitem{Marciano:1980pb}
W.~J. Marciano and A.~Sirlin, \emph{Radiative corrections to neutrino induced
  neutral current phenomena in the {SU(2)$_L$ x U(1)} theory},
  \href{https://doi.org/10.1103/PhysRevD.31.213,
  10.1103/PhysRevD.22.2695}{\emph{Phys. Rev.} {\bfseries D22} (1980) 2695}.

\bibitem{Awramik:2004ge}
M.~Awramik, M.~Czakon, A.~Freitas and G.~Weiglein, \emph{{Complete two-loop
  electroweak fermionic corrections to $\sin^{2} \theta^{\rm lept}_{\rm eff}$
  and indirect determination of the { Higgs} boson mass}},
  \href{https://doi.org/10.1103/PhysRevLett.93.201805}{\emph{Phys. Rev. Lett.}
  {\bfseries 93} (2004) 201805},
  [\href{https://arxiv.org/abs/hep-ph/0407317}{{\ttfamily hep-ph/0407317}}].

\bibitem{Hollik:2005va}
W.~Hollik, U.~Meier and S.~Uccirati, \emph{{The effective electroweak mixing
  angle $\sin^2\theta^{\mathrm{eff}}$ with two-loop fermionic contributions}},
  \href{https://doi.org/10.1016/j.nuclphysb.2005.10.015}{\emph{Nucl. Phys.}
  {\bfseries B731} (2005) 213--224},
  [\href{https://arxiv.org/abs/hep-ph/0507158}{{\ttfamily hep-ph/0507158}}].

\bibitem{Awramik:2006ar}
M.~Awramik, M.~Czakon and A.~Freitas, \emph{{Bosonic corrections to the
  effective weak mixing angle at $O(\alpha^2)$}},
  \href{https://doi.org/10.1016/j.physletb.2006.07.035}{\emph{Phys. Lett.}
  {\bfseries B642} (2006) 563--566},
  [\href{https://arxiv.org/abs/hep-ph/0605339}{{\ttfamily hep-ph/0605339}}].

\bibitem{Hollik:2006ma}
W.~Hollik, U.~Meier and S.~Uccirati, \emph{{The effective electroweak mixing
  angle $\sin^2\theta^{\mathrm{eff}}$ with two-loop bosonic contributions}},
  \href{https://doi.org/10.1016/j.nuclphysb.2006.12.001}{\emph{Nucl. Phys.}
  {\bfseries B765} (2007) 154--165},
  [\href{https://arxiv.org/abs/hep-ph/0610312}{{\ttfamily hep-ph/0610312}}].

\bibitem{Awramik:2008gi}
M.~Awramik, M.~Czakon, A.~Freitas and B.~Kniehl, \emph{{Two-loop electroweak
  fermionic corrections to $\sin^2 \theta^{\rm b \bar b}_{\rm eff}$}},
  \href{https://doi.org/10.1016/j.nuclphysb.2008.12.031}{\emph{Nucl. Phys.}
  {\bfseries B813} (2009) 174--187},
  [\href{https://arxiv.org/abs/0811.1364}{{\ttfamily 0811.1364}}].

\bibitem{Dubovyk:2016aqv}
I.~Dubovyk, A.~Freitas, J.~Gluza, T.~Riemann and J.~Usovitsch, \emph{{The
  two-loop electroweak bosonic corrections to $\sin^2\theta_{\rm eff}^{\rm
  b}$}}, \href{https://doi.org/10.1016/j.physletb.2016.09.012}{\emph{Phys.
  Lett.} {\bfseries B762} (2016) 184--189},
  [\href{https://arxiv.org/abs/1607.08375}{{\ttfamily 1607.08375}}].

\bibitem{Djouadi:1987gn}
A.~Djouadi and C.~Verzegnassi, \emph{Virtual very heavy top effects in
  {LEP/SLC} precision measurements},
  \href{https://doi.org/10.1016/0370-2693(87)91206-8}{\emph{Phys. Lett.}
  {\bfseries B195} (1987) 265--271}.

\bibitem{Djouadi:1987di}
A.~Djouadi, \emph{{O($\alpha \alpha_s$)} vacuum polarization functions of the
  standard model gauge bosons},
  \href{https://doi.org/10.1007/BF02812964}{\emph{Nuovo Cim.} {\bfseries A100}
  (1988) 357}.

\bibitem{Kniehl:1989yc}
B.~A. Kniehl, \emph{Two loop corrections to the vacuum polarizations in
  perturbative {QCD}},
  \href{https://doi.org/10.1016/0550-3213(90)90552-O}{\emph{Nucl. Phys.}
  {\bfseries B347} (1990) 86--104}.

\bibitem{Kniehl:1991gu}
B.~A. Kniehl and A.~Sirlin, \emph{{Dispersion relations for vacuum polarization
  functions in electroweak physics}},
  \href{https://doi.org/10.1016/0550-3213(92)90232-Z}{\emph{Nucl. Phys.}
  {\bfseries B371} (1992) 141--148}.

\bibitem{Djouadi:1993ss}
A.~Djouadi and P.~Gambino, \emph{{Electroweak gauge boson selfenergies:
  Complete QCD corrections}}, \href{https://doi.org/10.1103/PhysRevD.49.3499,
  10.1103/PhysRevD.53.4111}{\emph{Phys. Rev.} {\bfseries D49} (1994)
  3499--3511}, [\href{https://arxiv.org/abs/hep-ph/9309298}{{\ttfamily
  hep-ph/9309298}}].

\bibitem{Avdeev:1994db}
L.~Avdeev, J.~Fleischer, S.~Mikhailov and O.~Tarasov, \emph{{$0(\alpha
  \alpha_s^2)$ correction to the electroweak $\rho$ parameter}},
  \href{https://doi.org/10.1016/0370-2693(94)90573-8,
  10.1016/0370-2693(95)00269-Q}{\emph{Phys. Lett.} {\bfseries B336} (1994)
  560--566}, [\href{https://arxiv.org/abs/hep-ph/9406363}{{\ttfamily
  hep-ph/9406363}}].

\bibitem{Chetyrkin:1995ix}
K.~Chetyrkin, J.~H. K{\"u}hn and M.~Steinhauser, \emph{{Corrections of order
  ${O}(G_F M_t^2 \alpha_s^2)$ to the $\rho$ parameter}},
  \href{https://doi.org/10.1016/0370-2693(95)00380-4}{\emph{Phys. Lett.}
  {\bfseries B351} (1995) 331--338,
  \href{http://dx.doi.org/10.1016/0370--2693(95)00380--4}{doi:10.1016/0370--2693(95)00380--4}},
  [\href{https://arxiv.org/abs/hep-ph/9502291}{{\ttfamily hep-ph/9502291}}].

\bibitem{vanderBij:2000cg}
J.~J. van~der Bij, K.~G. Chetyrkin, M.~Faisst, G.~Jikia and T.~Seidensticker,
  \emph{{Three loop leading top mass contributions to the $\rho$ parameter}},
  \href{https://doi.org/10.1016/S0370-2693(01)00002-8}{\emph{Phys. Lett.}
  {\bfseries B498} (2001) 156--162},
  [\href{https://arxiv.org/abs/hep-ph/0011373}{{\ttfamily hep-ph/0011373}}].

\bibitem{Faisst:2003px}
M.~Faisst, J.~H. K{\"u}hn, T.~Seidensticker and O.~Veretin, \emph{{Three loop
  top quark contributions to the $\rho$ parameter}},
  \href{https://doi.org/10.1016/S0550-3213(03)00450-4}{\emph{Nucl. Phys.}
  {\bfseries B665} (2003) 649--662},
  [\href{https://arxiv.org/abs/hep-ph/0302275}{{\ttfamily hep-ph/0302275}}].

\bibitem{Schroder:2005db}
Y.~Schr{\"o}der and M.~Steinhauser, \emph{{Four-loop singlet contribution to
  the $\rho$ parameter}},
  \href{https://doi.org/10.1016/j.physletb.2005.06.085}{\emph{Phys. Lett.}
  {\bfseries B622} (2005) 124--130},
  [\href{https://arxiv.org/abs/hep-ph/0504055}{{\ttfamily hep-ph/0504055}}].

\bibitem{Chetyrkin:2006bj}
K.~G. Chetyrkin, M.~Faisst, J.~H. K{\"u}hn, P.~Maierh{\"o}fer and C.~Sturm,
  \emph{{Four-loop QCD corrections to the $\rho$ parameter}},
  \href{https://doi.org/10.1103/PhysRevLett.97.102003}{\emph{Phys. Rev. Lett.}
  {\bfseries 97} (2006) 102003},
  [\href{https://arxiv.org/abs/hep-ph/0605201}{{\ttfamily hep-ph/0605201}}].

\bibitem{Boughezal:2006xk}
R.~Boughezal and M.~Czakon, \emph{{Single scale tadpoles and $O(G_F m_t^2
  \alpha_s^3)$ corrections to the $\rho$ parameter}},
  \href{https://doi.org/10.1016/j.nuclphysb.2006.08.007}{\emph{Nucl. Phys.}
  {\bfseries B755} (2006) 221--238},
  [\href{https://arxiv.org/abs/hep-ph/0606232}{{\ttfamily hep-ph/0606232}}].

\bibitem{Chen:2020xzx}
L.~Chen and A.~Freitas, \emph{{Leading fermionic three-loop corrections to
  electroweak precision observables}},
  \href{https://doi.org/10.1007/JHEP07(2020)210}{\emph{JHEP} {\bfseries 07}
  (2020) 210}, [\href{https://arxiv.org/abs/2002.05845}{{\ttfamily
  2002.05845}}].

\bibitem{Chen:2020xot}
L.~Chen and A.~Freitas, \emph{{Mixed EW-QCD leading fermionic three-loop
  corrections at $\mathcal{O}(\alpha_s\alpha^2)$ to electroweak precision
  observables}}, \href{https://doi.org/10.1007/JHEP03(2021)215}{\emph{JHEP}
  {\bfseries 03} (2021) 215},
  [\href{https://arxiv.org/abs/2012.08605}{{\ttfamily 2012.08605}}].

\bibitem{Harlander:1997zb}
R.~Harlander, T.~Seidensticker and M.~Steinhauser, \emph{{Complete corrections
  of order $\alpha \alpha_s$ to the decay of the Z boson into bottom quarks}},
  \href{https://doi.org/10.1016/S0370-2693(98)00220-2}{\emph{Phys. Lett.}
  {\bfseries B426} (1998) 125--132},
  [\href{https://arxiv.org/abs/hep-ph/9712228}{{\ttfamily hep-ph/9712228}}].

\bibitem{Fleischer:1992fq}
J.~Fleischer, O.~Tarasov, F.~Jegerlehner and P.~Raczka, \emph{{Two loop $O
  (\alpha_s G_{\mu} m_t^2)$ corrections to the partial decay width of the $Z^0$
  into $b {\bar b}$ final states in the large top mass limit}},
  \href{https://doi.org/10.1016/0370-2693(92)90909-N}{\emph{Phys. Lett.}
  {\bfseries B293} (1992) 437--444}.

\bibitem{Buchalla:1992zm}
G.~Buchalla and A.~J. Buras, \emph{{QCD corrections to the $\bar s d Z$ vertex
  for arbitrary top quark mass}},
  \href{https://doi.org/10.1016/0550-3213(93)90110-B}{\emph{Nucl. Phys.}
  {\bfseries B398} (1993) 285--300}.

\bibitem{Degrassi:1993ij}
G.~Degrassi, \emph{{Current algebra approach to heavy top effects in $Z \to b +
  {\bar b}$}}, \href{https://doi.org/10.1016/0550-3213(93)90058-W}{\emph{Nucl.
  Phys.} {\bfseries B407} (1993) 271--289},
  [\href{https://arxiv.org/abs/hep-ph/9302288}{{\ttfamily hep-ph/9302288}}].

\bibitem{Chetyrkin:1993jp}
K.~Chetyrkin, A.~Kwiatkowski and M.~Steinhauser, \emph{Leading top mass
  corrections of order ${O} (\alpha \alpha_{s} m_t^2/m_{W}^2)$ to partial decay
  rate ${\Gamma} ({Z} \to b {\bar b})$},
  \href{https://doi.org/10.1142/S0217732393003172}{\emph{Mod. Phys. Lett.}
  {\bfseries A8} (1993) 2785--2792}.

\bibitem{Freitas:2013dpa}
A.~Freitas, \emph{{Two-loop fermionic electroweak corrections to the Z-boson
  width and production rate}},
  \href{https://doi.org/10.1016/j.physletb.2014.01.017}{\emph{Phys. Lett.}
  {\bfseries B730} (2014) 50--52},
  [\href{https://arxiv.org/abs/1310.2256}{{\ttfamily 1310.2256}}].

\bibitem{Dubovyk:2018rlg}
I.~Dubovyk, A.~Freitas, J.~Gluza, T.~Riemann and J.~Usovitsch, \emph{{Complete
  electroweak two-loop corrections to $Z$ boson production and decay}},
  \href{https://doi.org/10.1016/j.physletb.2018.06.037}{\emph{Phys. Lett.}
  {\bfseries B783} (2018) 86--94},
  [\href{https://arxiv.org/abs/1804.10236}{{\ttfamily 1804.10236}}].

\bibitem{Dubovyk:2019szj}
I.~Dubovyk, A.~Freitas, J.~Gluza, T.~Riemann and J.~Usovitsch,
  \emph{{Electroweak pseudo-observables and Z-boson form factors at two-loop
  accuracy}}, \href{https://doi.org/10.1007/JHEP08(2019)113}{\emph{JHEP}
  {\bfseries 08} (2019) 113},
  [\href{https://arxiv.org/abs/1906.08815}{{\ttfamily 1906.08815}}].

\bibitem{dizet645}
``{ZFITTER 6.42/DIZET 6.45}.'' \url{http://sanc.jinr.ru/users/zfitter/}.

\bibitem{Jadach:2022mbe}
S.~Jadach, B.~F.~L. Ward, Z.~Wa\u0327s, S.~A. Yost and A.~Siodmok,
  \emph{{Multi-photon Monte Carlo event generator KKMCee for lepton and quark
  pair production in lepton colliders}},
  \href{https://doi.org/10.1016/j.cpc.2022.108556}{\emph{Comput. Phys. Commun.}
  {\bfseries 283} (2023) 108556},
  [\href{https://arxiv.org/abs/2204.11949}{{\ttfamily 2204.11949}}].

\end{thebibliography}\endgroup

\end{document}